\documentclass{article}
\usepackage{graphicx}
\usepackage{amsmath}
\usepackage{amssymb}
\usepackage{amsfonts}
\usepackage{authblk}

\title{Background-Equivariant BRST Observables and i-Particle Propagators from an Auxiliary Quartet in SU(3) Yang-Mills}

\author[1]{M. M. Amaral\thanks{marcelo@gaugefreedom.com}}
\author[2]{V. E. R. Lemes\thanks{verlemes@gmail.com}}

\affil[1]{\small \em Gauge Freedom, Inc., A California Benefit Corporation, Los Angeles, CA, USA}
\affil[2]{\small \em Instituto de F\'\i sica, Universidade do Estado do Rio de Janeiro, \protect\\ Rua S\~{a}o Francisco Xavier 524, Maracan\~{a}, Rio de Janeiro - RJ, 20550-013, Brazil}

\date{}

\begin{document}

\maketitle

\begingroup
\renewcommand\thefootnote{}\footnotetext{Published in Nucl.\ Phys.\ B \textbf{1032} (2026) 117681, \texttt{doi:10.1016/j.nuclphysb.2026.117681}. Published under the CC BY 4.0 license (SCOAP$^3$).}
\addtocounter{footnote}{-1}
\endgroup

\begin{abstract}
In this work, we construct a BRST-exact quartet mechanism in $SU(3)$ Yang-Mills theory in the Landau gauge. The quartet sector is cohomologically trivial in the standard vacuum, ensuring equivalence to pure Yang-Mills theory. The transformation rules carry both commutator and anticommutator structures, enlarging the field content from eight to nine degrees of freedom.

The quartet prepotential is oriented by an auxiliary adjoint field $\mathcal N_3$ of mass dimension two --- the Cartan spurion --- through a BRST-exact term $-i\,s\,\mathrm{Tr}(\overline{\omega}\varphi\mathcal N_3)$, and the spurion sector is closed by a perfect-square potential $\lambda_1\,\mathrm{Tr}[(\mathcal N_3-\mu^2T^3)^2]$, whose constant seed $\mu^2T^3$ is the unique external dimension-two scale of the model. Being an algebraic quantum field, integrated over in the path integral, $\mathcal N_3$ carries a standard covariant BRST law $s\mathcal N_3=ig[\mathcal N_3,c]$, and its dimension-two assignment keeps the counterterm space finite by power counting, while the Slavnov--Taylor identity and the quantum action principle constrain it to appear only in covariant combinations. The coupled classical field equations of the quartet and spurion sectors admit an exact stationary solution, $\langle\mathcal N_3\rangle=\mu^2_{\mathrm{eff}}\,T^3$ with $\mu^2_{\mathrm{eff}}=[2\lambda_1\lambda/(2\lambda_1\lambda+1)]\,\mu^2$ and $\langle\overline{\varphi}\rangle\langle\varphi\rangle=(i\mu^2_{\mathrm{eff}}/\lambda)\,T^3$; integrating out the auxiliary field exactly reproduces the frozen-spurion formulation with a shifted effective coupling. Expanding around this saddle, the gauge fields acquire a mass matrix proportional to $d^{3ab}$ only, reproducing the $i$-particle propagator structure of earlier replica models without explicit breaking terms: a massless $(1,2)$ sector and complex-conjugate poles in the $(4,5)$, $(6,7)$ and diagonalized $(3,8)$ sectors. To respect the BRST doublet theorem, we separate background generation from observable cohomology. Introducing a background-equivariant covariant Cartan frame, we show the filtered $i$-particle bilinear is the lowest perturbative component of an all-orders off-shell BRST cocycle. At leading one-loop order, the two-point function of this lowest component retains a K\"all\'en--Lehmann representation with a real positive threshold and positive spectral density, despite the complex poles of the elementary propagators. The fully quantized action provides a consistent starting point for algebraic renormalization, establishing a systematic mechanism for recovering $i$-particle propagators and identifying BRST-controlled composite observables from a BRST-exact quartet extended to $SU(3)$.
\end{abstract}

\section{Introduction}
\label{sec:intro}
The problem of gluon confinement has been, for many years, one of the central issues of QCD. We know that the infrared behavior of the gluon field differs drastically from that of a particle excitation. One approach to obtaining information about the infrared behavior of gauge fields is the study of Gribov-type propagators. These propagators exhibit a complex pole structure that violates positivity, rendering the usual particle interpretation for the elementary gauge field inconsistent \cite{Gribov:1977wm,Zwanziger:1991ac,Zwanziger:1992qr,Dudal:2008sp}. Therefore, one must look for composite operators that carry physical meaning. The most standard proposal in the literature is that suitable composite operators admit a K\"all\'en--Lehmann representation with a positive spectral density and an associated real positive mass \cite{Baulieu:2009ha,Sorella:2010it}.

Several models have been proposed with the aim of obtaining a structure that exhibits both open Gribov-type propagators in their complex pole structure and facilitates the determination of physical operators. In particular, due to the similarity in terms of Green's functions, we highlight the work in \cite{Amaral:2013ReplicaSU3}, which studies a Yang–Mills model with $SU(3)$ breaking in the Landau gauge. In that model, a soft dimension-two breaking proportional to $d^{abc}A_\mu^bA_\mu^c$ induces a replica-like structure, allowing for the algebraic construction of observable candidates that exhibit K\"all\'en--Lehmann spectral representations with positive spectral density. Other models presenting such observables have also been introduced and adequately characterized algebraically by means of filtered sectors of BRST symmetry \cite{Amaral:2020Complex,Amaral:2023Observables}.

Our approach to the problem consists of introducing an auxiliary quartet sector that is cohomologically trivial in the standard vacuum. Consequently, the starting action does not alter the physics of the problem and remains completely equivalent to pure Yang-Mills theory. The quartet prepotential is oriented by an auxiliary adjoint field $\mathcal N_3$ of mass dimension two, integrated over in the path integral but devoid of a kinetic term: it is an algebraic quantum variable, and its covariant BRST law $s\mathcal N_3=ig[\mathcal N_3,c]$ is an ordinary symmetry assignment, free of any paradox about external fields transforming under the ghost. The spurion sector is closed by a perfect-square potential centered on a constant seed $\mu^2T^3$, the unique external dimension-two scale of the construction; with the standard spurion bookkeeping for the seed, the formal action remains BRST invariant, and the physical frame is reached by freezing the seed to its constant value. Because the orienting term is BRST-exact, the nilpotency and the Slavnov--Taylor identity of the theory are preserved exactly; the Cartan orientation and the mass scale are not imposed by hand but emerge from the coupled classical field equations of the quartet and spurion sectors, which admit an exact stationary solution around which the gauge fields acquire the $i$-particle mass structure. The dimension-two assignment is essential: it bounds the possible counterterms to at most quadratic order in the spurion, keeping the renormalization problem finite. To avoid conflicts with the standard BRST doublet theorem, we separate the generation of this background from the definition of the observables. Physical sectors can be captured by introducing a background-equivariant Cartan frame, which allows us to lift the filtered BRST symmetries of the gauge fields to an all-orders off-shell cocycle that exhibits a K\"all\'en--Lehmann representation.

The work is organized as follows: In Section~\ref{sec:color3}, we provide a brief review of the work in \cite{Amaral:2013ReplicaSU3}, detailing the complex pole propagators ($i$-particles) and the algebraic presentation of possible observables in the K\"all\'en--Lehmann sense. In Section~\ref{sec:quartet}, we introduce the concept of the BRST quartet, its $SU(3)$ extension, the auxiliary Cartan spurion field and the spurion-oriented BRST-exact prepotential. Section~\ref{sec:action} presents the fully quantized action, incorporating a Landau-type gauge fixing and source terms coupled to the symmetries of the fields, including a source for the spurion variation, and discusses the power counting of the spurion sector. Section~\ref{sec:background} is devoted to the coupled classical field equations of the quartet and spurion sectors, their exact Cartan-oriented solution, the reality conditions of the construction, and the derivation of the effective gauge-field mass matrix. Sections~\ref{sec:brst_bg} and \ref{sec:decomposition} detail the algebraic decomposition and filtering of the BRST operator. In Section~\ref{sec:full_lift}, we introduce the covariant Cartan frame and construct the full BRST lift of the filtered diagonal channel, proving its off-shell invariance. Section~\ref{sec:spectral} then computes the one-loop correlation function of the lowest component of this operator, demonstrating its positive K\"all\'en--Lehmann spectral representation. Finally, conclusions are presented in Section~\ref{sec:conclusion}.

\section{The original source model revisited}
\label{sec:color3}

We begin by recalling the special case studied in \cite{Amaral:2013ReplicaSU3}, namely the physical source aligned along the color-3 direction,
\begin{equation}
\langle J^a\rangle=m^2\delta^{a3},
\qquad m^2>0.
\label{eq:J-color3}
\end{equation}
At quadratic order, the gauge-field sector, assuming a Landau gauge, is
\begin{equation}
S_{\mathrm{quad}}
=
\frac12\int d^4x\,A_\mu^a(-\partial^2)A_\mu^a
+
\frac{i}{2}\int d^4x\,\langle J^a\rangle d^{abc}A_\mu^bA_\mu^c.
\label{eq:Squad-color3}
\end{equation}
Using the relevant symmetric coefficients
\begin{equation}
d^{338}=\frac{1}{\sqrt{3}},
\qquad
d^{344}=d^{355}=\frac12,
\qquad
d^{366}=d^{377}=-\frac12,
\label{eq:relevant-d-color3}
\end{equation}
the quadratic action decomposes into three distinct sectors. We denote by
\begin{equation}
\theta_{\mu\nu}(k)=\delta_{\mu\nu}-\frac{k_\mu k_\nu}{k^2}
\label{eq:transverse-projector}
\end{equation}
the transverse projector in momentum space.

The $(1,2)$ sector remains massless,
\begin{equation}
\langle A_\mu^1(k)A_\nu^1(-k)\rangle
=
\langle A_\mu^2(k)A_\nu^2(-k)\rangle
=
\frac{1}{k^2}\,\theta_{\mu\nu}(k),
\label{eq:prop12-color3}
\end{equation}
while the $(4,5)$ and $(6,7)$ sectors acquire opposite imaginary mass shifts,
\begin{equation}
\langle A_\mu^4(k)A_\nu^4(-k)\rangle
=
\langle A_\mu^5(k)A_\nu^5(-k)\rangle
=
\frac{1}{k^2+i\,m^2/2}\,\theta_{\mu\nu}(k),
\label{eq:prop45-color3}
\end{equation}
\begin{equation}
\langle A_\mu^6(k)A_\nu^6(-k)\rangle
=
\langle A_\mu^7(k)A_\nu^7(-k)\rangle
=
\frac{1}{k^2-i\,m^2/2}\,\theta_{\mu\nu}(k).
\label{eq:prop67-color3}
\end{equation}
Thus the two mixed off-diagonal subsectors form a conjugate pair of the standard $i$-particle type.

The diagonal $(3,8)$ sector is mixed already at quadratic level. Its action is
\begin{equation}
S_{(3,8)}
=
\int d^4x
\left\{
\frac12 A_\mu^3(-\partial^2)A_\mu^3
+
\frac12 A_\mu^8(-\partial^2)A_\mu^8
+
i\frac{m^2}{\sqrt{3}}A_\mu^3A_\mu^8
\right\},
\label{eq:S38-color3}
\end{equation}
which yields
\begin{equation}
\langle A_\mu^3(k)A_\nu^3(-k)\rangle
=
\langle A_\mu^8(k)A_\nu^8(-k)\rangle
=
\frac{k^2}{k^4+m^4/3}\,\theta_{\mu\nu}(k),
\label{eq:prop33-88-color3}
\end{equation}
\begin{equation}
\langle A_\mu^3(k)A_\nu^8(-k)\rangle
=
\langle A_\mu^8(k)A_\nu^3(-k)\rangle
=
-\,i\,\frac{m^2/\sqrt{3}}{k^4+m^4/3}\,\theta_{\mu\nu}(k).
\label{eq:prop38-color3}
\end{equation}
Introducing the rotated fields
\begin{equation}
U_\mu=\frac{1}{\sqrt{2}}(A_\mu^3+A_\mu^8),
\qquad
V_\mu=\frac{1}{\sqrt{2}}(-A_\mu^3+A_\mu^8),
\label{eq:UV-color3}
\end{equation}
the diagonal block takes the form
\begin{equation}
S_{(3,8)}
=
\frac12\int d^4x
\left[
U_\mu\left(-\partial^2+i\frac{m^2}{\sqrt{3}}\right)U_\mu
+
V_\mu\left(-\partial^2-i\frac{m^2}{\sqrt{3}}\right)V_\mu
\right].
\label{eq:S38diag-color3}
\end{equation}
Hence the $(3,8)$ sector also reorganizes into a conjugate $i$-particle pair, now with gap $m^2/\sqrt{3}$.

The important point for what follows is that the color-3 model supports two distinct neutral one-loop channels with acceptable analytic behavior. For the off-diagonal sectors, we define
\begin{equation}
E_{\mu\nu}^{3\pm}
=
\frac{1}{\sqrt{2}}\left(F_{\mu\nu}^4\pm iF_{\mu\nu}^5\right),
\qquad
E_{\mu\nu}^{2\pm}
=
\frac{1}{\sqrt{2}}\left(F_{\mu\nu}^6\pm iF_{\mu\nu}^7\right).
\label{eq:Epm-color3}
\end{equation}
Under the residual Cartan generator $h_1=T^3=\lambda^3/2$, these obey
\begin{equation}
[h_1,E^{2\pm}]=\mp\frac12 E^{2\pm},
\qquad
[h_1,E^{3\pm}]=\pm\frac12 E^{3\pm}.
\label{eq:charges-color3}
\end{equation}
so that the bilinear
\begin{equation}
\phi
=
E_{\mu\nu}^{2+}E_{\mu\nu}^{3+}
+
E_{\mu\nu}^{3-}E_{\mu\nu}^{2-}
=
F_{\mu\nu}^4F_{\mu\nu}^6-F_{\mu\nu}^5F_{\mu\nu}^7
\label{eq:phi-color3}
\end{equation}
is neutral and Hermitian. Because it is built from the conjugate pair carried by the $(4,5)$ and $(6,7)$ propagators, its one-loop two-point function admits the standard K\"all\'en--Lehmann construction \cite{Baulieu:2009ha,Sorella:2010it}.

The second channel comes from the diagonalized $(3,8)$ sector and is described by
\begin{equation}
\chi
=
(\partial_\mu U_\nu-\partial_\nu U_\mu)
(\partial_\mu V_\nu-\partial_\nu V_\mu).
\label{eq:chi-color3}
\end{equation}
Since $U_\mu$ and $V_\mu$ also form a conjugate pair, the corresponding one-loop correlator again has the standard $i$-particle structure.

Accordingly, the one-loop thresholds of the two physical channels are
\begin{equation}
\tau_\phi=m^2,
\qquad
\tau_\chi=\frac{2m^2}{\sqrt{3}}.
\label{eq:thresholds-color3}
\end{equation}
The original color-3 model is therefore characterized by the simultaneous coexistence of a replica-like off-diagonal physical channel and a diagonal physical channel.
An important discussion is in order here, which forms the core motivation for this work. The replica model cited above relies on an explicit soft breaking of gauge invariance. Therefore, although one can evaluate condensed states in the K\"all\'en--Lehmann sense at the leading order, there is no underlying mechanism guaranteeing that this construction is radiatively stable or systematically extendable. Only a fully quantized, BRST-exact action can provide a consistent algebraic framework to generate and compute such observables. This points directly to the need for an auxiliary mechanism—such as a BRST-exact quartet—that can dynamically induce this precise $i$-particle background without explicitly breaking the fundamental gauge symmetry.

\section{Auxiliary Quartet}
\label{sec:quartet}
We first recall the standard BRST-quartet mechanism. Consider a pair of
bosonic fields $(\overline{\varphi},\varphi)$ and a pair of fermionic
fields $(\overline{\omega},\omega)$, with BRST transformations
\[
s\overline{\omega}=\overline{\varphi},\qquad
s\overline{\varphi}=0,\qquad
s\varphi=\omega,\qquad
s\omega=0.
\]
The action
\begin{equation}
S_{q}=\int d^{4}x \left[
\partial_{\mu}\overline{\varphi}\partial^{\mu}\varphi
-\partial_{\mu}\overline{\omega}\partial^{\mu}\omega
- m^{2} (\overline{\varphi}\varphi-\overline{\omega}\omega)
-\lambda(\overline{\varphi}\varphi-\overline{\omega}\omega)^2
\right]
\end{equation}
is BRST-exact:
\begin{equation} 
S_{q}= s \int d^{4}x \left[
\partial_{\mu}\overline{\omega}\partial^{\mu}\varphi
- m^{2}\overline{\omega}\varphi
- \lambda(\overline{\omega}\varphi)
(\overline{\varphi}\varphi-\overline{\omega}\omega)
\right].
\end{equation}
Therefore this sector is cohomologically trivial: it does not enlarge
the local BRST cohomology of physical observables. In BRST-invariant
quantities, the bosonic and fermionic quartet contributions cancel in
the usual doublet-sector sense, modulo the standard assumptions on the
regularization and functional measure.

As usual, the BRST operator carries a ghost charge, which means that for the fields $\overline{\varphi}$ and $\varphi$ to be ordinary scalars, the field $\overline{\omega}$ must have ghost charge $-1$ and the field $\omega$ must have ghost charge $1$. This type of structure has already been used in other works \cite{deSa:2020rnu} and is, although little remembered in the literature, a property already demonstrated in \cite{Piguet:1995er}.

However, we cannot use it in its simplest form. It is necessary to introduce an SU(N) group, in the case N=3, such that the fields take values over the group and interact with gauge fields. Thus, let us recall that the Yang-Mills action is:
\begin{equation}
S_{YM}= \int d^{4}x  \operatorname{Tr} \left\{\frac{1}{4}(F_{\mu\nu}F^{\mu\nu})\right\}, \,\,\, F_{\mu\nu}=\partial_{\mu}A_{\nu}-\partial_{\nu}A_{\mu}-ig[ A_{\mu},A_{\nu}]
\end{equation}
where the trace is understood over the group generators (with $N^2-1$ components). The field is written as $A_{\mu}=T^{a}A^{a}_{\mu}$.
The corresponding BRST transformations are:
\begin{equation}
sA_{\mu}= - D_{\mu}c = -(\partial_{\mu}c -ig[A_{\mu},c]),\,\,\, sc=-\frac{i}{2}g\{c,c\} = -igc^{2}.
\end{equation}

Although the Yang-Mills action is expected to give rise to condensed states of the glueball type, obtaining a closed form for these candidates is not trivial, and the calculation of such states in the general case is not well defined, despite existing attempts in this direction, such as in \cite{Capri:2010pg}.

\subsection{Extending the quartet to an $SU(N)$ group and a slightly different transformation}

Now we not only extend the quartet transformation to $SU(N)$ but also introduce a ``matter sector'' into the transformation that has a curious property: it carries not only the commutator but also the anticommutator.
\begin{eqnarray}
s\overline{\omega}&=& \overline{\varphi} -igc\overline{\omega},\nonumber \\
s\overline{\varphi}&=& -ig c\overline{\varphi},\nonumber \\
s\varphi &=& \omega +ig\varphi c,\nonumber \\
s\omega&=& -ig\omega c.
\label{eq:brst-quartet}
\end{eqnarray}
It is important to emphasize two things here. First, it remains a strictly nilpotent BRST quartet ($s^2 = 0$ on all fields). The relative minus sign in the $s\omega$ transformation ensures nilpotency under the standard graded Leibniz rule. To clarify the structure before proceeding, the quantum numbers and transformation properties of the fields are summarized in Table \ref{tab:brst_charges}.

\begin{table}[h]
\centering
\begin{tabular}{|c|c|c|c|c|}
\hline
\textbf{Field} & \textbf{Type} & \textbf{UV dim} & \textbf{Ghost Number} & \textbf{BRST} $s\Phi$ \\
\hline
$A_\mu$ & Commuting & $1$ & $0$ & $-D_\mu c$ \\
$c$ & Anticommuting & $0$ & $1$ & $-igc^2$ \\
$\overline{\varphi}$ & Commuting & $1$ & $0$ & $-igc\overline{\varphi}$ \\
$\varphi$ & Commuting & $1$ & $0$ & $\omega + ig\varphi c$ \\
$\overline{\omega}$ & Anticommuting & $1$ & $-1$ & $\overline{\varphi} - igc\overline{\omega}$ \\
$\omega$ & Anticommuting & $1$ & $1$ & $-ig\omega c$ \\
$\mathcal N_3$ & Commuting (auxiliary) & $2$ & $0$ & $ig[\mathcal N_3,c]$ \\
\hline
\end{tabular}
\caption{Field content, gradings, and BRST transformations for the $SU(3)$ extended quartet, including the dimension-two auxiliary Cartan spurion field $\mathcal N_3$ (algebraic, non-propagating, integrated over in the path integral).}
\label{tab:brst_charges}
\end{table}

Throughout, $s$ acts as an odd graded derivation, $s(XY) = (sX)Y + (-1)^{|X|}X(sY)$, and traces over products containing Grassmann-valued matrices are understood with graded cyclicity. With these conventions, the gauge-covariant quartet action is strictly $s$-exact. 

Second, the "matter sector" of the transformation has the property of also carrying the anticommutator, which implies that these fields are not 8 components in $SU(3)$ but rather 9 components, with the component 0 associated to the identity matrix present in the anticommutator:
\begin{eqnarray}
\varphi &=& \varphi^{0} T^{0} + \varphi^{a} T^{a}, \,\,\,\,T^{0} = \frac{1}{\sqrt{6}} \mathbf{1}_{3}\nonumber \\
\omega &=& \omega^{0} T^{0} + \omega^{a} T^{a}.
\label{eq:nine-components}
\end{eqnarray}
The choice of $T^{0}$ with this normalization is to facilitate the trace calculation, so that $\operatorname{Tr}(T^AT^B)=\tfrac12\delta^{AB}$ for $A,B\in\{0,1,\dots,8\}$. The gauge and ghost fields, in contrast, have only the eight $SU(3)$ components, $A_\mu=A_\mu^aT^a$, $c=c^aT^a$, $a=1,\dots,8$, and the same holds for the spurion field, $\mathcal N_3=\mathcal N_3^aT^a$, since its transformation law \eqref{eq:spurion-brst} below involves only the commutator with the ghost.

The covariant derivatives of the fields are defined as:
\begin{eqnarray}
D_{\mu} \overline{\omega} &=& \partial_{\mu} \overline{\omega} - ig A_\mu \overline{\omega},\qquad
D_{\mu} \overline{\varphi} = \partial_{\mu} \overline{\varphi} - ig A_\mu \overline{\varphi},\nonumber \\
D_{\mu} \varphi &=& \partial_{\mu} \varphi + ig \varphi A_\mu ,\qquad
D_{\mu} \omega = \partial_{\mu} \omega + ig \omega A_\mu ,
\end{eqnarray}
such that the derivatives obey:
\begin{eqnarray}
sD_{\mu} \overline{\omega}&=& D_{\mu} \overline{\varphi} -igcD_{\mu} \overline{\omega},\nonumber \\
sD_{\mu} \overline{\varphi}&=& -igc D_{\mu} \overline{\varphi},\nonumber \\
sD_{\mu} \varphi &=& D_{\mu} \omega +ig(D_{\mu} \varphi )c,\nonumber \\
sD_{\mu} \omega&=& -ig(D_{\mu} \omega )c.
\end{eqnarray}

We emphasize that, by construction, $\overline{\varphi}$ and $\varphi$ are \emph{independent complex fields}, not Hermitian conjugates: they obey different transformation laws (left multiplication for $\overline{\varphi}$, right multiplication for $\varphi$). The reality conditions and the path-integral contour appropriate to this complexification are discussed in Section~\ref{subsec:reality}.

\subsection{The auxiliary Cartan spurion field}
\label{subsec:spurion}

We enlarge the field content by an auxiliary adjoint field $\mathcal N_3=\mathcal N_3^aT^a$ ($a=1,\dots,8$) --- the Cartan spurion --- commuting, with ghost number zero and \emph{mass dimension two}, transforming as
\begin{equation}
s\mathcal N_3 = ig[\mathcal N_3,c],
\label{eq:spurion-brst}
\end{equation}
which is strictly nilpotent, as follows from $s^2c=0$ and the Jacobi identity. Three features of this assignment are essential for the consistency of the whole construction, and we stress them from the outset.

\noindent\textbf{(i) Auxiliary quantum character.} The field $\mathcal N_3$ is a bona fide quantum variable: it is integrated over in the path integral. It is auxiliary because it has no kinetic term --- a field of mass dimension two admits no renormalizable kinetic term, since $\operatorname{Tr}(D_\mu\mathcal N_3D^\mu\mathcal N_3)$ already has dimension six --- and it enters the action at most quadratically, so that its functional integral is Gaussian and can be performed exactly (see Section~\ref{subsec:eom}). This resolves at the outset any conceptual objection to the transformation law \eqref{eq:spurion-brst}: a purely external, inert background should indeed not transform under the ghost, but $\mathcal N_3$ is not external --- it is a quantum field, and Eq.~\eqref{eq:spurion-brst} is an ordinary covariant symmetry assignment to a quantum variable, of the same nature as the adjoint transformation of any matter field. In fact, the transformation law is not optional: since the combination $Y=\overline{\varphi}\varphi-\overline{\omega}\omega$ transforms in the adjoint, $sY=-ig[c,Y]$, BRST invariance of the orienting term $\operatorname{Tr}(Y\mathcal N_3)$ \emph{requires} $s\mathcal N_3=ig[\mathcal N_3,c]$. An inert $\mathcal N_3$ would make the action non-invariant, returning the construction to the uncontrolled status of an explicit breaking.

\noindent\textbf{(ii) The seed and the perfect-square potential.} The spurion sector is closed by the potential
\begin{equation}
S_{\mathrm{spurion}}=\lambda_1\int d^{4}x\,\operatorname{Tr}\big[\big(\mathcal N_3-\mu^2T^3\big)^2\big],
\qquad \lambda_1>0,
\label{eq:spurion-potential}
\end{equation}
where the constant $\mu^2T^3$, of mass dimension two, is the \emph{unique external scale} (the seed) of the model. Two complementary points of view clarify its status. Formally, assigning the seed itself the adjoint spurion law $s(\mu^2T^3)=ig\mu^2[T^3,c]$ --- the standard spurion bookkeeping of algebraic renormalization \cite{Piguet:1995er} --- makes the combination $Z=\mathcal N_3-\mu^2T^3$ covariant, $sZ=ig[Z,c]$, and the potential \eqref{eq:spurion-potential} exactly BRST invariant. In the physical frame, where the seed is frozen to its constant value, the formal invariance is expressed as a Ward identity with a classical breaking that is \emph{linear in the quantum fields}, $s\,\Sigma\propto ig\lambda_1\mu^2\int d^4x\,\operatorname{Tr}\big([T^3,\mathcal N_3]\,c\big)$; linear breakings of this type are stable under radiative corrections, their coefficients being protected by the standard algebraic results on softly broken symmetries \cite{Piguet:1995er}. The potential \eqref{eq:spurion-potential} is what stabilizes the spurion: without it, the coupled field equations of Section~\ref{sec:background} would admit only the trivial stationary point, and the $i$-particle background would not be generated.

\noindent\textbf{(iii) Dimension two and power counting.} A dimensionless spurion with a dimensionless vacuum value would be radiatively unstable in an uncontrollable way: since arbitrary powers of a dimension-zero quantity have dimension zero, counterterms involving arbitrarily high powers of the spurion would be allowed, and no finite set of counterterms could renormalize the theory. Assigning the spurion mass dimension two closes this loophole: locality and power counting bound any integrated counterterm to be at most \emph{quadratic} in $\mathcal N_3$. The counterterm space of the spurion sector is therefore finite, and will be displayed explicitly in Section~\ref{subsec:powercounting}. Moreover, since $\mathcal N_3$ transforms as an adjoint field under the BRST operator, the theory admits the covariant derivative
\begin{equation}
D_\mu\mathcal N_3=\partial_\mu\mathcal N_3-ig[A_\mu,\mathcal N_3],
\qquad
sD_\mu\mathcal N_3=ig[D_\mu\mathcal N_3,c],
\label{eq:spurion-covder}
\end{equation}
whose covariance follows directly from the nilpotency of $s$ and the Jacobi identity; it enters no kinetic term, but it belongs to the covariant building blocks $(F_{\mu\nu},D_\mu\mathcal N_3,D_\mu F_{\mu\nu},\dots)$ on which the local BRST algebra closes. The Slavnov--Taylor identity, preserved at the quantum level by the quantum action principle, then constrains every counterterm to be an invariant polynomial of these covariant blocks: the spurion can only appear through covariant combinations, and only in the forms compatible with its dimension-two assignment. This is the precise sense in which a symmetry --- the Slavnov--Taylor identity itself --- fixes the spurion to appear in the action only in the form displayed below.

The key identity for what follows is the BRST variation
\begin{equation}
s\,\operatorname{Tr}\big(\overline{\omega}\varphi\mathcal N_3\big)
=
\operatorname{Tr}\big[(\overline{\varphi}\varphi-\overline{\omega}\omega)\mathcal N_3\big],
\label{eq:spurion-exact}
\end{equation}
in which all ghost-dependent terms cancel exactly by the graded cyclicity of the trace: the terms $-ig\operatorname{Tr}(c\overline{\omega}\varphi\mathcal N_3)$ and $-ig\operatorname{Tr}(\overline{\omega}\varphi\mathcal N_3 c)$ differ by a minus sign under cyclic permutation, while the terms involving $[\mathcal N_3,c]$ cancel among themselves. The right-hand side of Eq.~\eqref{eq:spurion-exact} contains no explicit ghost field.

\subsection{The spurion-oriented quartet action}
\label{subsec:quartet-action}

We can now define the quartet action. Instead of the trace mass term of the standard quartet, we orient the prepotential along the covariant spurion:
\begin{equation}
\begin{split}
S_{qg}
={}&
s\int d^{4}x \operatorname{Tr}\bigg\{
D_{\mu}\overline{\omega}D^{\mu}\varphi
+ \frac{\lambda}{2}(\overline{\omega}\varphi)(\overline{\varphi}\varphi-\overline{\omega}\omega)
- i\,\overline{\omega}\varphi\mathcal N_3
\bigg\}
\\
&+\lambda_1\int d^{4}x\,\operatorname{Tr}\big[\big(\mathcal N_3-\mu^2T^3\big)^2\big],
\end{split}
\label{eq:Sqg-prepotential}
\end{equation}
which, using Eq.~\eqref{eq:spurion-exact}, evaluates to
\begin{equation}
\begin{split}
S_{qg}=\int d^{4}x \operatorname{Tr}\bigg\{&
D_{\mu}\overline{\varphi}D^{\mu}\varphi
- D_{\mu}\overline{\omega}D^{\mu}\omega
+ \frac{\lambda}{2}(\overline{\varphi}\varphi - \overline{\omega}\omega)^{2}
- i(\overline{\varphi}\varphi - \overline{\omega}\omega)\mathcal N_3
\\
&+\lambda_1\big(\mathcal N_3-\mu^2T^3\big)^2
\bigg\}.
\end{split}
\label{eq:Sqg}
\end{equation}
The $s$-exact part is manifestly invariant under the full nilpotent BRST operator, including the spurion transformation law \eqref{eq:spurion-brst}, and is therefore cohomologically trivial; the perfect-square spurion potential is invariant in the formal (seed-covariant) sense discussed in item (ii) of Section~\ref{subsec:spurion}. Each term has dimension four: the couplings $\lambda$ and $\lambda_1$ are dimensionless, the spurion carries the dimension two of the orienting interaction, and the seed $\mu^2T^3$ sets the mass scale.

Redefining the Yang-Mills action as:
\begin{eqnarray}
S_{New} &=& \int d^{4}x \; \left\{ \operatorname{Tr}\left( \frac{1}{4}F_{\mu\nu}F^{\mu\nu} \right) + \operatorname{Tr}\left( D_{\mu}\overline{\varphi}D^{\mu}\varphi - D_{\mu}\overline{\omega}D^{\mu}\omega \right) \right. \nonumber \\
&& \quad\left. -\, i \operatorname{Tr}\left[ \left( \overline{\varphi}\varphi - \overline{\omega}\omega \right)\mathcal N_3 \right] + \frac{\lambda}{2} \operatorname{Tr}\left[ \left( \overline{\varphi}\varphi - \overline{\omega}\omega \right)^{2} \right] \right. \nonumber \\
&& \quad\left. +\, \lambda_1\operatorname{Tr}\big[\big(\mathcal N_3-\mu^2T^3\big)^2\big] \right\}
\label{eq:SNew}
\end{eqnarray}
we see that, from the cohomological point of view, we do not alter the Yang-Mills action at all, as far as the characterization of observables is concerned: the quartet sector is $s$-exact, and the spurion sector, being built from the algebraic auxiliary $\mathcal N_3$, does not enlarge the local BRST cohomology either; the standard doublet theorem \cite{Piguet:1995er} applies. A similar, albeit weaker, mechanism occurs in the usual gauge fixing, where the ghost sector, written using the $b^{a}$ field which forms a doublet with the antighost $\overline{c}^{a}$, allows one to obtain that physical observables are independent of the ghost sector.

Two remarks are in order. First, the Euclidean action \eqref{eq:SNew} is complex, since the orienting term carries an explicit factor of $i$; this is the same analytic situation as in the replica model of Section~\ref{sec:color3}, whose soft-breaking term is also purely imaginary, and the physical content will be extracted through composite correlators with real K\"all\'en--Lehmann representations. Second, in the trivial background (vanishing scalar expectation values and the spurion at its zero stationary point), the quartet-plus-spurion term leaves the physical cohomology exactly equivalent to that of pure Yang-Mills. By evaluating the source-free BRST-exact theory on the nontrivial Cartan-oriented stationary point derived in Section~\ref{sec:background}, we effectively probe a different physical sector. The BRST symmetry reflects this background shift, altering the local cohomology such that certain field combinations carry information that is not observed in the trivial vacuum \cite{Piguet:1995er}. The corresponding physical interpretation is then formally secured in the background-equivariant extended complex introduced in Section~\ref{sec:full_lift}.

\section{Fully Quantized Action}
\label{sec:action}
To establish a framework where explicit calculations—including renormalization—can be performed, a gauge-fixing procedure is necessary. We adopt the Landau gauge, for which the gauge-fixing action takes the form:
\begin{equation}
S_{gf}=\int d^{4}x \operatorname{Tr} \{ ib\partial_{\mu}A^{\mu} + \overline{c}\partial_{\mu}D^{\mu}c\}
\end{equation}
with the nilpotent BRST transformations for the gauge and ghost fields given by:
\begin{eqnarray}
sA_{\mu}&=& - D_{\mu}c = -(\partial_{\mu}c -ig[A_{\mu},c]),\nonumber \\ 
sc&=& -igc^{2},\nonumber \\
s\overline{c}&=&ib,\,\,\, sb=0.
\end{eqnarray}
This allows the gauge-fixing term to be written as a BRST variation:
\begin{equation}
S_{gf}=s\int d^{4}x \operatorname{Tr} \{\overline{c}\partial_{\mu}A^{\mu}\}.
\end{equation}

Because the BRST transformations of the fields involve composite operators (products of fields at the same point), external sources must be introduced to properly define these non-linear variations. This step is strictly necessary to assemble the "fully quantized" action required for algebraic renormalization. Although a complete proof of renormalizability to all orders falls outside the scope of this article, we present the corresponding source action for completeness:
\begin{eqnarray}
S_{f}&=&\int d^{4}x \operatorname{Tr} \{ -\Omega_{\mu}D^{\mu}c -igLc^{2} \nonumber \\ 
&+& (\overline{\varphi} -igc\overline{\omega})Y_{\bar{\omega}} 
- igc\overline{\varphi}Y_{\bar{\varphi}} 
+ Y_{\varphi}(\omega+ ig\varphi c) 
- ig Y_{\omega}\omega c
+ ig\,\Upsilon[\mathcal N_3, c] \},
\label{eq:Sf}
\end{eqnarray}
where $\Omega_\mu, L, Y_{\bar{\omega}}, Y_{\bar{\varphi}}, Y_{\varphi},$ and $Y_{\omega}$ act as the external sources coupled to the respective BRST variations, and the new external source $\Upsilon$ is coupled to the BRST variation of the spurion, $s\mathcal N_3=ig[\mathcal N_3,c]$. Coupling this variation to a source is the standard way to control symmetry sectors with nonlinear transformations in the algebraic renormalization framework \cite{Piguet:1995er}: it allows the spurion to be subjected to the same power-counting and Ward-identity analysis as the remaining fields, and it makes the covariance of the construction under the adjoint rotation of the background manifest at the quantum level. The same construction applies verbatim to a second spurion whenever it is included in the action sector.

The total action is then:
\begin{equation}
S = S_{New} + S_{gf},
\end{equation}
and the fully quantized action is:
\begin{equation}
\Sigma = S + S_{f}.
\end{equation}

The dimensions and ghost numbers of the external sources are fixed by
requiring each source term in $S_f$ to have dimension four and ghost
number zero:
\[
\begin{array}{c|c|c}
\text{Source} & \text{Dimension} & \text{Ghost number} \\
\hline
\Omega_\mu & 3 & -1 \\
L & 4 & -2 \\
Y_{\bar\omega} & 3 & 0 \\
Y_{\bar\varphi} & 3 & -1 \\
Y_\varphi & 3 & -1 \\
Y_\omega & 3 & -2 \\
\Upsilon & 2 & -1
\end{array}
\]

With these sources, the complete action satisfies the Slavnov-Taylor
identity
\[
\mathcal S(\Sigma)=0,
\]
where
\[
\begin{aligned}
\mathcal S(\Sigma)=
\int d^4x\,\operatorname{Tr}\Bigg(
&\frac{\delta \Sigma}{\delta \Omega_\mu}
\frac{\delta \Sigma}{\delta A_\mu}
+
\frac{\delta \Sigma}{\delta L}
\frac{\delta \Sigma}{\delta c}
+
ib\frac{\delta \Sigma}{\delta \bar c}
\\
&+
\frac{\delta \Sigma}{\delta Y_{\bar\omega}}
\frac{\delta \Sigma}{\delta \bar\omega}
+
\frac{\delta \Sigma}{\delta Y_{\bar\varphi}}
\frac{\delta \Sigma}{\delta \bar\varphi}
+
\frac{\delta \Sigma}{\delta Y_{\varphi}}
\frac{\delta \Sigma}{\delta \varphi}
+
\frac{\delta \Sigma}{\delta Y_{\omega}}
\frac{\delta \Sigma}{\delta \omega}
\\
&+
\frac{\delta \Sigma}{\delta \Upsilon}
\frac{\delta \Sigma}{\delta \mathcal N_3}
+
2\lambda_{1}\mu^{2}T^{3}
\frac{\delta \Sigma}{\delta \Upsilon}
\Bigg).
\end{aligned}
\]
up to the conventional left/right functional-derivative signs required
for fermionic fields. The last term, $2\lambda_{1}\mu^{2}T^{3}\,\delta\Sigma/\delta\Upsilon$, reflects the fact that the seed $\mu^{2}T^{3}$ is kept as a constant external matrix: it compensates the non-vanishing BRST variation of the linear spurion coupling contained in the perfect-square potential, so that the identity holds exactly for the action as defined.

The fully quantized action defined above in the Landau gauge is the natural choice for any future loop calculations or algebraic proofs. Crucially, the choice of the Landau condition ($\partial_\mu A^\mu = 0$) plays a fundamental role in the quadratic reduction of the theory. When the quartet fields are shifted around a nontrivial background (as will be done in the subsequent section), potential gauge-scalar mixing terms of the form $A_\mu \partial^\mu \tilde{\varphi}$ become proportional to $\partial_\mu A^\mu$ after integration by parts, and thus vanish. This ensures the ghost decoupling properties and guarantees that the physical gauge field propagators cleanly decouple from the longitudinal scalar fluctuations.

\subsection{Power counting and stability of the spurion sector}
\label{subsec:powercounting}

We now make precise the stability properties announced in Section~\ref{subsec:spurion}. Because the spurion has mass dimension two, ghost number zero, and enters the extended Slavnov--Taylor identity through the source $\Upsilon$, the quantum action principle restricts the possible local counterterms to invariant polynomials of the covariant building blocks $(F_{\mu\nu},D_\mu\mathcal N_3,D_\mu F_{\mu\nu},\overline{\varphi},\varphi,\overline{\omega},\omega,\text{ghosts},\text{sources})$ of dimension at most four. The spurion dependence is consequently bounded to at most quadratic order, and the complete list of $\mathcal N_3$-dependent invariant counterterms is finite:

\begin{enumerate}
\item \emph{Quadratic spurion term} $\operatorname{Tr}(\mathcal N_3^2)$: this is the tree-level perfect-square structure itself, Eq.~\eqref{eq:spurion-potential}; its renormalization amounts to a renormalization of the dimensionless coupling $\lambda_1$. A term $(\operatorname{Tr}\mathcal N_3)^2$ would also be allowed by quantum numbers, but vanishes identically because $\mathcal N_3=\mathcal N_3^aT^a$ is traceless.

\item \emph{Seed-dependent terms} $\mu^2\operatorname{Tr}(T^3\mathcal N_3)$ and the constant $\mu^4\operatorname{Tr}[(T^3)^2]$: these are precisely the terms generated by expanding the perfect square \eqref{eq:spurion-potential}; the first is the tree-level tadpole that drives the spurion to its nontrivial stationary value, and the second contributes only to the vacuum energy. At the tree level their coefficients are tied to $\lambda_1$ and to the seed; whether this tie persists under renormalization --- that is, whether the perfect-square combination is maintained --- is a stability question to be settled by the explicit counterterm analysis. The essential point here is that the setup reduces that problem to the finite, controlled set of candidates listed above.

\item \emph{Linear mixing} of the form $\operatorname{Tr}\big[\mathcal N_3(\overline{\varphi}\varphi-\overline{\omega}\omega)\big]$: this is the tree-level orienting structure itself, Eq.~\eqref{eq:spurion-exact}, which is $s$-exact; its renormalization amounts to a renormalization of the parameters of the model, without affecting the cohomology.

\item \emph{Cubic and higher powers of $\mathcal N_3$}: forbidden by power counting, since $\dim\mathcal N_3=2$. This is the crucial point: no infinite tower of spurion counterterms can arise.
\end{enumerate}

In particular, a gauge-sector mass counterterm $\operatorname{Tr}(\mathcal N_3 A_\mu A_\mu)$, although allowed by dimension, is not invariant under the Slavnov--Taylor identity — its BRST variation does not vanish — and is therefore excluded. The tree-level structure $d^{3ab}$ induced by the background is thus the unique dimension-two deformation of the gauge sector compatible with the symmetries of the fully quantized action. This is the precise content of the radiative-stability statement for the pole structure: it follows from power counting plus the Slavnov--Taylor identity, while the complete counterterm and operator-mixing analysis remains a well-defined, finite algebraic problem left for future work.

\section{Quartet Background and Effective Mass Matrix}
\label{sec:background}

\noindent
The quartet sector is BRST-exact and therefore cohomologically trivial in the 
standard vacuum. Evaluating the theory in a nontrivial background for the scalar 
fields does not break the BRST symmetry of the action; the operator $s$ remains strictly 
nilpotent and the action remains $s$-invariant. However, when expanded around this 
background, the explicit form of the BRST transformations acquires additional 
terms linear in the background values. Consequently, the \emph{local BRST 
cohomology} changes relative to the trivial vacuum: new $s$-closed classes 
become possible, while some previously trivial classes may become nontrivial. 
This is the algebraic mechanism that allows physical observables — in particular, 
composite operators with a K\"all\'en--Lehmann representation — to emerge in the 
cohomology, even though such operators are absent in the symmetric phase. 

\subsection{Coupled classical field equations and their exact Cartan-oriented solution}
\label{subsec:eom}

With vanishing fermionic expectation values $\langle\overline{\omega}\rangle=\langle\omega\rangle=0$, and writing $Y\equiv\overline{\varphi}\varphi-\overline{\omega}\omega$ with $\langle Y\rangle=\langle\overline{\varphi}\rangle\langle\varphi\rangle$, the bosonic potential of the quartet-plus-spurion sector reads
\begin{equation}
V[\overline{\varphi},\varphi,\mathcal N_3]
=
\operatorname{Tr}\left[
\frac{\lambda}{2}\,Y^2
-
i\,Y\mathcal N_3
+
\lambda_1\big(\mathcal N_3-\mu^2T^3\big)^2
\right].
\label{eq:potential}
\end{equation}
The potential is a sum of quadratic forms in $Y$ and $\mathcal N_3$; consequently, the stationarity conditions are \emph{linear} matrix equations, and the stationary point can be found exactly. Varying the potential with respect to the independent complex fields $\overline{\varphi}$, $\varphi$ and to the auxiliary spurion $\mathcal N_3$ gives the full classical field equations
\begin{equation}
\varphi\left(\lambda\,Y-i\,\mathcal N_3\right)=0,
\qquad
\left(\lambda\,Y-i\,\mathcal N_3\right)\overline{\varphi}=0,
\qquad
-\,iY+2\lambda_1\big(\mathcal N_3-\mu^2T^3\big)=0.
\label{eq:eom}
\end{equation}
Stationarity is a \emph{matrix} condition, not merely a condition on a trace. For full-rank factors $\overline{\varphi},\varphi$, the first two equations impose
\begin{equation}
\lambda\,Y=i\,\mathcal N_3,
\label{eq:eom-Y}
\end{equation}
while the spurion equation gives
\begin{equation}
i\,Y=2\lambda_1\big(\mathcal N_3-\mu^2T^3\big).
\label{eq:eom-N}
\end{equation}
Combining Eqs.~\eqref{eq:eom-Y} and \eqref{eq:eom-N}, and assuming $2\lambda_1\lambda+1\neq0$ (automatic for positive couplings), one obtains the exact, unique full-rank stationary solution on the branch \(\lambda Y=i\mathcal N_3\), purely along the Cartan direction of the seed:
\begin{equation}
\boxed{\;\langle\mathcal N_3\rangle=\mu^2_{\mathrm{eff}}\,T^3,
\qquad
X\equiv\langle\overline{\varphi}\rangle\langle\varphi\rangle=\frac{i\mu^2_{\mathrm{eff}}}{\lambda}\,T^3,
\qquad
\mu^2_{\mathrm{eff}}\equiv\frac{2\lambda_1\lambda}{2\lambda_1\lambda+1}\,\mu^2\;}
\label{eq:eom-solution}
\end{equation}
for any factorization of $X$ into full-rank factors; for instance,
\begin{equation}
\langle\overline{\varphi}\rangle=v\,\mathbf 1,
\qquad
\langle\varphi\rangle=\frac{i\mu^2_{\mathrm{eff}}}{\lambda v}\,T^3,
\qquad v\neq 0.
\label{eq:vev-representative}
\end{equation}
The field equations fix only the product $X$; the split between the two factors is a flat direction of the vacuum manifold, and all physical quantities below depend only on the product. A convenient representative, which we adopt from now on, is
\begin{equation}
\langle\overline{\varphi}\rangle=\frac{a}{\sqrt3}\,\mathbf 1,
\qquad
\langle\varphi\rangle=i\sqrt6\,a\,T^3,
\qquad
\frac{\mu^2_{\mathrm{eff}}}{\lambda}=\sqrt2\,a^2.
\label{eq:vev-choice}
\end{equation}

Five remarks are in order.

\noindent\textbf{(i) Saddle character.} Besides the solution manifold \eqref{eq:eom-solution}, Eqs.~\eqref{eq:eom} admit degenerate rank-deficient stationary points (for instance $\varphi=0$ with $T^3\overline{\varphi}=0$). Since the Euclidean potential is complex, stationarity — not global minimization — is the meaningful criterion, and we expand around the prescribed saddle \eqref{eq:eom-solution}. This is the sense in which the background is chosen among the solutions of the classical field equations: it is a genuine stationary point of the fully quantized, BRST-exact action, so the quadratic propagators obtained below are vacuum propagators of the shifted theory.

\noindent\textbf{(ii) Exact integration of the auxiliary field.} Because $\mathcal N_3$ has no kinetic term and appears only quadratically in \eqref{eq:potential}, its functional integral is Gaussian and can be performed exactly. Completing the square,
\begin{equation}
\lambda_1\operatorname{Tr}\big[\big(\mathcal N_3-\mu^2T^3-\tfrac{i}{2\lambda_1}Y\big)^2\big]
-i\mu^2\operatorname{Tr}\big(YT^3\big)
+\frac{1}{4\lambda_1}\operatorname{Tr}\big(Y^2\big),
\label{eq:gaussian-square}
\end{equation}
and discarding the field-independent Gaussian determinant (a constant, absorbed into the normalization), the spurion integration produces the effective quartet potential
\begin{equation}
V_{\mathrm{eff}}[Y]
=
\frac{\lambda_{\mathrm{eff}}}{2}\operatorname{Tr}\big(Y^2\big)
-
i\mu^2\operatorname{Tr}\big(YT^3\big),
\qquad
\lambda_{\mathrm{eff}}=\lambda+\frac{1}{2\lambda_1},
\label{eq:potential-effective}
\end{equation}
whose stationary point $\langle Y\rangle=(i\mu^2/\lambda_{\mathrm{eff}})T^3=(i\mu^2_{\mathrm{eff}}/\lambda)T^3$ coincides exactly with Eq.~\eqref{eq:eom-solution}. This shows explicitly that the auxiliary field introduces no additional dynamics: the present formulation is the quadratic resolution of the frozen-spurion formulation, with the effective coupling $\lambda_{\mathrm{eff}}$ and the physical scale $\mu^2_{\mathrm{eff}}$.

\noindent\textbf{(iii) Trace condition.} The solution satisfies $\operatorname{Tr}X=0$: the product is purely traceless. No trace condition of the type $\operatorname{Tr}(\langle\overline{\varphi}\rangle\langle\varphi\rangle)=m^2/\lambda$ is needed nor imposed; the mass scale is fixed by the matrix equations \eqref{eq:eom} themselves, through the seed of the dimension-two spurion potential.

\noindent\textbf{(iv) Trace component of $\langle\overline{\varphi}\rangle$.} The representative \eqref{eq:vev-choice} places the VEV of $\overline{\varphi}$ along the identity. This component does not couple linearly to the gauge fields, since $\operatorname{Tr}(T^0T^a)=0$ for $a=1,\dots,8$ and there is no ghost associated with the $U(1)$ direction in the $SU(3)$ BRST complex. Its only role is to parametrize the factorization of the product \eqref{eq:eom-solution}; any other factorization is equally admissible and physically equivalent.

\noindent\textbf{(v) The role of the seed.} If the seed is removed, $\mu^2\to0$, the unique full-rank solution collapses to the trivial point $\langle\mathcal N_3\rangle=\langle Y\rangle=0$: the perfect-square potential \eqref{eq:spurion-potential} is what stabilizes the nontrivial Cartan-oriented vacuum, and the seed $\mu^2T^3$ is the unique dimension-two external parameter from which the whole mass scale of the construction descends, through the combination $\mu^2_{\mathrm{eff}}$.

\subsection{Reality conditions and the BRST status of the vacuum}
\label{subsec:reality}

As stressed in Section~\ref{sec:quartet}, $\overline{\varphi}$ and $\varphi$ are independent complex fields with distinct (left/right) transformation laws; they are \emph{not} Hermitian conjugates. Accordingly:

\noindent\textbf{(i)} The vacuum expectation values \eqref{eq:vev-choice} obey $\langle\overline{\varphi}\rangle\neq\langle\varphi\rangle$, and $\langle\varphi\rangle$ is anti-Hermitian. There is no contradiction: no reality condition relating the two fields is ever imposed. The path-integral contour for the complexified scalar sector is chosen so that the expansion around the saddle \eqref{eq:eom-solution} is well defined, in the same spirit as the complex Euclidean actions of the replica model \cite{Amaral:2013ReplicaSU3} and of the $i$-particle toy models \cite{Baulieu:2009ha}. The Gaussian integral over the auxiliary spurion is algebraic and contour-independent, since its quadratic form $\lambda_1>0$ is positive. The physical output of the construction — the K\"all\'en--Lehmann representation of Section~\ref{sec:spectral} — is real and positive.

\noindent\textbf{(ii)} The BRST operator remains strictly nilpotent and the action strictly $s$-invariant after the shift; the Slavnov--Taylor identity of Section~\ref{sec:action} is unaffected by the background.

\noindent\textbf{(iii)} At the level of expectation values, $\langle s\varphi\rangle=\langle\omega\rangle+ig\langle\varphi c\rangle=0$ and $\langle s\overline{\varphi}\rangle=-ig\langle c\overline{\varphi}\rangle=0$, since $\langle c\rangle=\langle\omega\rangle=0$. On the other hand, $\langle s\overline{\omega}\rangle=\langle\overline{\varphi}\rangle\neq 0$ at tree level. We state explicitly what this means: the BRST symmetry of the \emph{action} is exact, while the BRST \emph{charge} is formally spontaneously broken at the vacuum — but only along the cohomologically trivial quartet doublet $(\overline{\omega},\overline{\varphi})$, whose excitations decouple from all physical (BRST-invariant) quantities by the standard doublet theorem \cite{Piguet:1995er}. The physical observables are defined as cocycles of the background-equivariant differential constructed in Section~\ref{sec:full_lift}, for which the quartet sector is inert. These three statements — exact symmetry of the action, prescribed saddle among the classical solutions, and formal breaking confined to the trivial quartet sector — are logically independent and are kept separate throughout.

\subsection{Generation of the mass term for the gauge fields}
\label{subsec:massterm}

The gauge-field mass term is generated by the kinetic term of the quartet scalars. From
\begin{equation}
\operatorname{Tr}\big(D_\mu\overline{\varphi}D^\mu\varphi\big)
\supset
g^2\operatorname{Tr}\big(A_\mu\overline{\varphi}\varphi A^\mu\big),
\end{equation}
the shift to the background \eqref{eq:eom-solution} gives, by cyclicity of the trace,
\begin{equation}
\mathcal L_{\mathrm{mass}}
=
g^2\operatorname{Tr}\big(X A_\mu A^\mu\big)
=
\frac{ig^2\mu^2_{\mathrm{eff}}}{\lambda}\operatorname{Tr}\big(T^3A_\mu A_\mu\big).
\label{eq:Lmass-step1}
\end{equation}
Since $A_\mu$ has only the eight $SU(3)$ components and
\begin{equation}
\operatorname{Tr}\big(T^3T^cT^d\big)=\frac14\big(d^{3cd}+if^{3cd}\big),
\end{equation}
with the $f$-term dropping out by the symmetry $c\leftrightarrow d$, one obtains
\begin{equation}
\mathcal L_{\mathrm{mass}}
=
\frac{ig^2\mu^2_{\mathrm{eff}}}{4\lambda}\,d^{3cd}A_\mu^cA_\mu^d
=
\frac{i}{2}M^{cd}A_\mu^cA_\mu^d,
\qquad
M^{cd}=\frac{g^2\mu^2_{\mathrm{eff}}}{2\lambda}\,d^{3cd}=\frac{g^2a^2}{\sqrt2}\,d^{3cd}.
\label{eq:Lmass-final}
\end{equation}
The mass matrix is \emph{purely} proportional to $d^{3ab}$: because the product $X$ contains no identity and no $T^8$ component, no $\delta^{cd}$ or $d^{8cd}$ structures are generated. In particular, the fact that the gauge field has no $T^0$ component does not by itself eliminate identity-type contributions — $\operatorname{Tr}(T^0T^cT^d)=\tfrac{1}{2\sqrt6}\delta^{cd}\neq 0$ — so the purity of \eqref{eq:Lmass-final} is a nontrivial property of the saddle \eqref{eq:eom-solution}, not of the trace structure alone.

The correct eigenvalues of $d^{3ab}$ in the adjoint representation of $SU(3)$ are:

\begin{center}
\begin{tabular}{|c|c|c|}
\hline
Subspace & Eigenvalue $\lambda_3$ & Multiplicity \\
\hline
$(1,2)$ sector & $0$ & 2 \\
$(3,8)$ sector & $\pm \dfrac{1}{\sqrt{3}}$ & 2 \\
$(4,5)$ sector & $\dfrac{1}{2}$ & 2 \\
$(6,7)$ sector & $-\dfrac{1}{2}$ & 2 \\
\hline
\end{tabular}
\end{center}

and therefore the proper eigenvalues of the effective mass matrix $M^{ab}$ are:
\[
\mu_{1,2} = 0, \qquad \mu_{3,8} = \pm \frac{g^2 a^{2}}{\sqrt{6}}, \qquad \mu_{4,5,6,7} = \pm \frac{g^2 a^{2}}{2\sqrt{2}}. 
\]

From this it is clear that the $i$-particle structure of the original model has been reproduced.

\subsection{Propagators of the gauge fields}

Defining
\[
m_1^2 = \frac{g^2 a^{2}}{\sqrt{6}}=\frac{g^2\mu^2_{\mathrm{eff}}}{2\sqrt3\,\lambda}, \qquad m_2^2 = \frac{g^2 a^{2}}{2\sqrt{2}}=\frac{g^2\mu^2_{\mathrm{eff}}}{4\lambda},
\]
the propagators in the original basis are:

\begin{equation}
\langle A^1 A^1 \rangle = \langle A^2 A^2 \rangle = \frac{1}{k^2}\,\theta_{\mu\nu},
\qquad
\langle A^1 A^2 \rangle = 0,
\end{equation}

\begin{equation}
\langle A^3 A^3 \rangle = \langle A^8 A^8 \rangle = \frac{k^2}{k^4 + m_1^4}\,\theta_{\mu\nu},
\qquad
\langle A^3 A^8 \rangle = \frac{-i m_1^2}{k^4 + m_1^4}\,\theta_{\mu\nu},
\end{equation}

\begin{equation}
\langle A^4 A^4 \rangle = \langle A^5 A^5 \rangle = \frac{1}{k^2 + i m_2^2}\,\theta_{\mu\nu},
\qquad
\langle A^4 A^5 \rangle = 0,
\end{equation}

\begin{equation}
\langle A^6 A^6 \rangle = \langle A^7 A^7 \rangle = \frac{1}{k^2 - i m_2^2}\,\theta_{\mu\nu},
\qquad
\langle A^6 A^7 \rangle = 0.
\end{equation}

\subsection{Field combinations that diagonalize the propagators}

For the off-diagonal charged sectors, we define the standard complex combinations:
\begin{equation}
A_\mu^{45\pm} = \frac{1}{\sqrt{2}}\left( A_\mu^4 \pm i A_\mu^5 \right),
\qquad
A_\mu^{67\pm} = \frac{1}{\sqrt{2}}\left( A_\mu^6 \pm i A_\mu^7 \right).
\end{equation}

For the diagonal $(3,8)$ sector, we rotate the fields into a real conjugate pair $U_\mu$ and $V_\mu$:
\begin{equation}
U_\mu = \frac{1}{\sqrt{2}}\left( A_\mu^3 + A_\mu^8 \right),
\qquad
V_\mu = \frac{1}{\sqrt{2}}\left( A_\mu^8 - A_\mu^3 \right).
\end{equation}

The diagonalized propagators are:

\begin{equation}
\begin{split}
\langle U_\mu(k) U_\nu(-k) \rangle &= \frac{1}{k^2 + i m_1^2}\,\theta_{\mu\nu}(k), \\
\langle V_\mu(k) V_\nu(-k) \rangle &= \frac{1}{k^2 - i m_1^2}\,\theta_{\mu\nu}(k), \\
\langle U_\mu(k) V_\nu(-k) \rangle &= 0,
\end{split}
\end{equation}

\begin{equation}
\langle A^{45+} A^{45-} \rangle = \langle A^{45-} A^{45+} \rangle = \frac{1}{k^2 + i m_2^2}\,\theta_{\mu\nu}(k),
\end{equation}

\begin{equation}
\langle A^{67+} A^{67-} \rangle = \langle A^{67-} A^{67+} \rangle = \frac{1}{k^2 - i m_2^2}\,\theta_{\mu\nu}(k),
\end{equation}

and all propagators with the same charge (e.g., $\langle A^{45+}A^{45+}\rangle$) vanish.

\subsection{Diagonalized curvatures}

The corresponding curvatures are redefined to match these diagonalized variables:

\begin{equation}
\begin{aligned}
F^{1+}_{\mu\nu} &= \frac{1}{\sqrt{2}} (F^1_{\mu\nu} + iF^2_{\mu\nu}), \qquad F^{1-}_{\mu\nu} = \frac{1}{\sqrt{2}} (F^1_{\mu\nu} - iF^2_{\mu\nu}), \\[4pt]
F^{U}_{\mu\nu} &= \frac{1}{\sqrt{2}}(F^3_{\mu\nu} + F^8_{\mu\nu}), \qquad F^{V}_{\mu\nu} = \frac{1}{\sqrt{2}}(F^8_{\mu\nu} - F^3_{\mu\nu}), \\[4pt]
F^{45+}_{\mu\nu} &= \frac{1}{\sqrt{2}}(F^4_{\mu\nu} + iF^5_{\mu\nu}), \qquad F^{45-}_{\mu\nu} = \frac{1}{\sqrt{2}}(F^4_{\mu\nu} - iF^5_{\mu\nu}), \\[4pt]
F^{67+}_{\mu\nu} &= \frac{1}{\sqrt{2}}(F^6_{\mu\nu} + iF^7_{\mu\nu}), \qquad F^{67-}_{\mu\nu} = \frac{1}{\sqrt{2}}(F^6_{\mu\nu} - iF^7_{\mu\nu}).
\end{aligned}
\end{equation}

\section{BRST Symmetry in the Nontrivial Background and Cohomology Structure}
\label{sec:brst_bg}

\subsection{Vacuum Choice}

The background values that orient the system, including a trace component for $\overline{\varphi}$, are chosen as the representative \eqref{eq:vev-choice} of the exact stationary manifold \eqref{eq:eom-solution}:

\begin{equation}
\langle \overline{\varphi} \rangle = \frac{a}{\sqrt{3}}\mathbf{1}, \qquad
\langle \varphi \rangle = i\sqrt{6}\,a\,T^3, \qquad a \in \mathbb{R},
\qquad
\langle\overline{\varphi}\rangle\langle\varphi\rangle = i\sqrt{2}\,a^2 T^3 = \frac{i\mu^2_{\mathrm{eff}}}{\lambda}T^3
\end{equation}

The generators of $SU(3)$ are:

\begin{equation}
T^3 = \frac{1}{2} \begin{pmatrix} 1 & 0 & 0 \\ 0 & -1 & 0 \\ 0 & 0 & 0 \end{pmatrix}, \qquad
T^8 = \frac{1}{2\sqrt{3}} \begin{pmatrix} 1 & 0 & 0 \\ 0 & 1 & 0 \\ 0 & 0 & -2 \end{pmatrix}
\end{equation}

The $U(1)$ (trace) generator is normalized as:
\begin{equation}
T^0 = \frac{1}{\sqrt{6}} \mathbf{1}_3
\end{equation}

\subsection{Original BRST Transformations}

The relevant full BRST operator $s$ is nilpotent ($s^2 = 0$) and acts on the gauge and scalar sectors as:

\begin{equation}
\begin{aligned}
s\overline{\varphi} &= -ig\, c\, \overline{\varphi}, \qquad & s\varphi &= \omega + ig\, \varphi\, c, \\
sA_\mu &= -\partial_\mu c + ig [A_\mu, c], \qquad & sc &= -ig\, c^2, \qquad s\bar{c} = ib, \qquad sb=0
\end{aligned}
\end{equation}

\subsection{Field Decomposition}

Expanding the scalar fields around their background values:

\begin{equation}
\begin{aligned}
\overline{\varphi} &= \frac{a}{\sqrt{3}}\mathbf{1} + \tilde{\overline{\varphi}}, \\
\varphi &= i\sqrt{6}\,a\,T^3 + \tilde{\varphi}, \\
A_\mu &= A_\mu^a T^a, \qquad c = c^a T^a
\end{aligned}
\end{equation}

The quantum fluctuations are decomposed in the basis $\{T^0, T^a\}$:

\begin{equation}
\tilde{\overline{\varphi}} = \tilde{\overline{\varphi}}^0 T^0 + \tilde{\overline{\varphi}}^a T^a, \qquad 
\tilde{\varphi} = \tilde{\varphi}^0 T^0 + \tilde{\varphi}^a T^a
\end{equation}

\subsection{BRST Transformations in the Nontrivial Background}

Substituting the decompositions and using the fact that the background satisfies $s\langle\varphi\rangle = s\langle\overline{\varphi}\rangle = 0$ at the level of expectation values (since $\langle c\rangle=\langle\omega\rangle=0$), we obtain:

\begin{equation}
\begin{aligned}
s\tilde{\overline{\varphi}} &= -ig\, c \left(\frac{a}{\sqrt{3}}\mathbf{1}\right) - ig\, c \tilde{\overline{\varphi}}, \\[4pt]
s\tilde{\varphi} &= \omega + ig \left(i\sqrt{6}\,a\,T^3\right) c + ig\, \tilde{\varphi} c, \\[4pt]
sA_\mu &= -\partial_\mu c + ig [A_\mu, c], \\[4pt]
sc &= -ig\, c^2, \qquad s\bar{c} = ib, \qquad sb = 0
\end{aligned}
\end{equation}

\subsection{The Operator $s_{0}$}

The operator $s_{0}$ is defined as the filtered, linearized part of $s$ \cite{Piguet:1995er}. Focusing on the scalar fluctuations and the gauge-ghost couplings, and using
\begin{equation}
T^3c=\frac{c^3}{6}\mathbf 1+\frac12\big(d^{3bc}+if^{3bc}\big)c^bT^c,
\end{equation}
we have:

\begin{equation}
\begin{aligned}
s_{0} A_\mu^a &= -\partial_\mu c^a, \\[4pt]
s_{0} \tilde{\overline{\varphi}}^0 &= 0, \\[4pt]
s_{0} \tilde{\overline{\varphi}}^a &= -\frac{ig\,a}{\sqrt3}\, c^a, \\[4pt]
s_{0} \tilde{\varphi}^0 &= \omega^0 - g\,a\,c^3, \\[4pt]
s_{0} \tilde{\varphi}^c &= \omega^c - \frac{g\sqrt6\,a}{2}\, (d^{3bc} + i f^{3bc})\, c^b, \\[4pt]
s_{0} c^a &= 0, \qquad s_{0} \bar{c}^a = ib^a, \qquad s_{0} b^a = 0
\end{aligned}
\end{equation}

\subsection{Decomposition $s = s_{0} + s_1$}

The full BRST operator decomposes as:

\begin{equation}
s = s_{0} + s_1
\end{equation}

where $s_1$ contains bilinear or higher terms in quantum fields. The nilpotency properties imply:

\begin{equation}
s_{0}^2 = 0, \qquad s_1^2 = 0, \qquad s_{0} s_1 + s_1 s_{0} = 0
\end{equation}

\subsection{Gauge-Ghost Canceling Combinations}

Two important combinations where the gauge-ghost ($c^a$) dependence exactly cancels under $s_{0}$ are:

\begin{equation}
\partial_\mu \tilde{\overline{\varphi}}^a - \frac{ig\,a}{\sqrt3}\, A_\mu^a
\qquad (a=1,\dots,8)
\end{equation}

\begin{equation}
\partial_\mu \tilde{\varphi}^0 - g\,a\, A_\mu^3
\end{equation}

Notice that the first set of combinations is strictly $s_{0}$-invariant, while the second transforms as $s_{0}\big(\partial_\mu \tilde{\varphi}^0 - g\,a\, A_\mu^3\big)=\partial_\mu \omega^0$, a pure quartet fluctuation which is exact under the full BRST operator ($\omega=s\varphi$). This explicitly demonstrates that the gauge-ghost degrees of freedom decouple, leaving behind only cohomologically trivial quartet fluctuations.

\subsection{Cohomology Structure}

The introduction of the nontrivial background does not break the BRST symmetry of the action; rather, it modifies the cohomology structure of the problem. According to standard algebraic renormalization theorems, the cohomology of the full operator $s$ is isomorphic to a subspace of the cohomology of the linearized operator $s_{0}$:

\begin{equation}
H(s) \hookrightarrow H(s_{0})
\end{equation}

That is, the physical states (cohomology of $s$) form a subset of the states that are $s_{0}$-closed modulo $s_{0}$-exact \cite{Piguet:1995er}.

For clarity, we distinguish explicitly between three complexes used in this work, which should not be conflated:

\begin{enumerate}
\item the \emph{full} local BRST cohomology $H(s)$ of the shifted theory, defined on local polynomials in all fields, sources and the spurion;
\item the \emph{linearized/filtered} cohomology $H(s_0)$, where $s_0$ is nilpotent on the whole local algebra by construction; the embedding $H(s)\hookrightarrow H(s_0)$ is the standard result quoted above;
\item the \emph{background-equivariant extended} cohomology of Section~\ref{sec:full_lift}, in which the algebra is enlarged by the covariant spurions with adjoint BRST law (the auxiliary spurion $\mathcal N_3$ in the action sector, and the dimensionless frame fields $\widehat{\mathcal N}_i$ of the observable sector).
\end{enumerate}

A reversible shift of the fields by a background does not by itself change the full local cohomology $H(s)$; what changes is the filtered complex $s_0$ and, later, the extended algebra including the spurions. The filtered Cartan operator $\delta_{(0)}$ introduced in Section~\ref{sec:decomposition} belongs to yet another bookkeeping device: it is nilpotent only after restriction to the ghost-free curvature-polynomial sector specified there, and all statements involving $H(\delta_{(0)})$ are to be understood within that restricted complex.

Thus, the lowest nontrivial component of any full $s$-cocycle must define a $s_{0}$-closed class. In this sense, $H(s_{0})$ provides the first rigorous approximation to the full BRST cohomology, while the complete physical representative must still be lifted to an $s$-closed expression. This is precisely the role of the background-equivariant construction in Section~\ref{sec:full_lift}: the quadratic $i$-particle operator is first identified at the filtered level and then completed into a full $s$-closed cocycle relative to the covariant background frame, ensuring independence from the trivial quartet fluctuations.

\section{Algebraic Decomposition and Filtering of the BRST Operator}
\label{sec:decomposition}

In this section, we present the decomposition of the BRST operator $s$ based on a filtering that counts the number of ghosts associated with the Cartan directions $c^3$ and $c^8$. Unlike the quantum fluctuation expansion discussed in Section~\ref{sec:brst_bg}, this filtering reveals a richer algebraic structure where $s$ decomposes into three non-trivial components. This specific decomposition is useful because it isolates the
filtration-preserving component $\delta_{(0)}$, which controls the
lowest Cartan-neutral part of the physical channel. Since
$\delta_{(0)}$ is not nilpotent by itself, this lowest component should
not be interpreted as an autonomous cohomology class. Its full
cohomological meaning is supplied by the background-equivariant BRST
lift constructed in Section~\ref{sec:full_lift}.

\subsection{Filtering Operator}

We define the number operator $N$:

\begin{equation}
N = c^3 \frac{\delta}{\delta c^3} + c^8 \frac{\delta}{\delta c^8}
\end{equation}

The eigenvalues of $N$ for the fields of the pure gauge sector (without scalars) are:

\begin{center}
\begin{tabular}{|c|c|}
\hline
\textbf{Field} & \textbf{Eigenvalue} $N$ \\
\hline
$A_\mu^1, A_\mu^2, A_\mu^4, A_\mu^5, A_\mu^6, A_\mu^7$ & $0$ \\
$A_\mu^3, A_\mu^8$ & $0$ \\
$c^1, c^2, c^4, c^5, c^6, c^7$ & $0$ \\
$c^3, c^8$ & $1$ \\
\hline
\end{tabular}
\end{center}

\subsection{Algebraic Decomposition of the BRST Operator}

The full BRST operator $s$ is nilpotent ($s^2 = 0$) and can be expanded in components with definite eigenvalue under the commutator with $N$:

\begin{equation}
s = \sum_{k} \delta_{(k)}, \qquad [N, \delta_{(k)}] = k \, \delta_{(k)}
\end{equation}

Since $s$ can increase, preserve or decrease the number of $c^3, c^8$, the decomposition contains three non-vanishing terms:

\begin{equation}
s =\delta_{(-1)} +\delta_{(0)} +\delta_{(1)}
\end{equation}

\subsection{Nilpotency Relations by Degree}

The condition $s^2 = 0$ implies, for each degree $m$:

\begin{equation}
\sum_{i+j=m} \delta_{(i)} \delta_{(j)} = 0
\end{equation}

Explicitly:
\begin{align}
\text{Degree } -2: &\quad\delta_{(-1)}^2 = 0 \label{eq:grau-2}\\
\text{Degree } -1: &\quad\delta_{(-1)}\delta_{(0)} +\delta_{(0)}\delta_{(-1)} = 0 \label{eq:grau-1}\\
\text{Degree } 0: &\quad\delta_{(0)}^2 +\delta_{(-1)}\delta_{(1)} +\delta_{(1)}\delta_{(-1)} = 0 \label{eq:grau0}\\
\text{Degree } 1: &\quad\delta_{(0)}\delta_{(1)} +\delta_{(1)}\delta_{(0)} = 0 \label{eq:grau1}\\
\text{Degree } 2: &\quad\delta_{(1)}^2 = 0 \label{eq:grau2}
\end{align}

\textbf{Important observation:} Unlike the linearized operator defined in Section~\ref{sec:brst_bg}, this filtered $\delta_{(0)}$ is \textbf{not} nilpotent by itself. Its nilpotency is satisfied only through the degree $0$ relation:

\begin{equation}
\delta_{(0)}^2 = - (\delta_{(-1)}\delta_{(1)} +\delta_{(1)}\delta_{(-1)})
\end{equation}

\subsection{Action of $\delta_{(-1)}$ (Decreases $N$ by 1)}

The operator $\delta_{(-1)}$ acts \textbf{only} on the ghosts $c^3$ and $c^8$:

\begin{equation}
\begin{aligned}
\delta_{(-1)} c^3 &= g \, c^1 c^2 + \frac{g}{2} c^4 c^5 - \frac{g}{2} c^6 c^7 \\[4pt]
\delta_{(-1)} c^8 &= \frac{\sqrt{3}g}{2} c^4 c^5 + \frac{\sqrt{3}g}{2} c^6 c^7
\end{aligned}
\end{equation}

For all other fields ($A_\mu^a$, $c^i$ with $i \in \{1,2,4,5,6,7\}$):

\begin{equation}
\delta_{(-1)} A_\mu^a = 0, \qquad\delta_{(-1)} c^i = 0
\end{equation}

\subsection{Action of $\delta_{(0)}$ (Preserves $N$)}

\subsubsection*{Diagonal Gauge Fields}
\begin{equation}
\begin{aligned}
\delta_{(0)} A_\mu^3 &= -g (A_\mu^1 c^2 - A_\mu^2 c^1) - \frac{g}{2} (A_\mu^4 c^5 - A_\mu^5 c^4) + \frac{g}{2} (A_\mu^6 c^7 - A_\mu^7 c^6) \\[4pt]
\delta_{(0)} A_\mu^8 &= - \frac{\sqrt{3}g}{2} (A_\mu^4 c^5 - A_\mu^5 c^4) - \frac{\sqrt{3}g}{2} (A_\mu^6 c^7 - A_\mu^7 c^6)
\end{aligned}
\end{equation}

For the gauge fields with $i \in I = \{1,2,4,5,6,7\}$:

\begin{equation}
\delta_{(0)} A_\mu^i = -\partial_\mu c^i - g \sum_{j \in I} (f^{i3j} A_\mu^3 + f^{i8j} A_\mu^8) c^j - g \sum_{j,k \in I} f^{ijk} A_\mu^j c^k
\end{equation}

\subsubsection*{Ghosts}

For the ghosts with $i \in I$:

\begin{equation}
\delta_{(0)} c^i = \frac{g}{2} \sum_{j,k \in I} f^{ijk} c^j c^k
\end{equation}

And for the ghosts $c^3$ and $c^8$:

\begin{equation}
\delta_{(0)} c^3 = 0, \qquad\delta_{(0)} c^8 = 0
\end{equation}

\subsection{Action of $\delta_{(1)}$ (Increases $N$ by 1)}

\subsubsection*{Diagonal Gauge Fields}
\begin{equation}
\delta_{(1)} A_\mu^3 = -\partial_\mu c^3, \qquad\delta_{(1)} A_\mu^8 = -\partial_\mu c^8
\end{equation}

For the gauge fields with $i \in I$:

\begin{equation}
\delta_{(1)} A_\mu^i = -g \sum_{j \in I} f^{ij3} A_\mu^j c^3 - g \sum_{j \in I} f^{ij8} A_\mu^j c^8
\end{equation}

\subsubsection*{Ghosts}

For $i \in I$:

\begin{equation}
\delta_{(1)} c^i = g \sum_{j \in I} f^{ij3} c^j c^3 + g \sum_{j \in I} f^{ij8} c^j c^8
\end{equation}

And for $c^3$ and $c^8$:

\begin{equation}
\delta_{(1)} c^3 = 0, \qquad\delta_{(1)} c^8 = 0
\end{equation}

\subsection{Restriction to the Curvature-Polynomial Sector and Nilpotency of the Restricted $\delta_{(0)}$}
\label{subsec:restricted-delta0}

As established in Eq.~\eqref{eq:grau0}, the filtration-preserving
component $\delta_{(0)}$ is not nilpotent on the full local field
algebra. Its failure to be nilpotent is controlled by the components
$\delta_{(-1)}$ and $\delta_{(1)}$, which change the number of
Cartan ghosts $c^3$ and $c^8$.

For the construction of physical composite insertions, however, we do
not consider ghost-dependent operators. We restrict the filtered
analysis to the ghost-number-zero curvature-polynomial sector relevant
for local physical observables. More precisely, we consider local
polynomials
\[
X=X(F_{\mu\nu},D_\rho F_{\mu\nu},\ldots)
\]
with no ghost insertions and with no active dependence on the Cartan
ghosts $c^3,c^8$. In this restricted sector the admissible filtered
candidates are taken to satisfy
\begin{equation}
\delta_{(-1)}X=0,
\qquad
\delta_{(1)}X=0.
\label{eq:restricted-sector-conditions}
\end{equation}
Equivalently, the sector is obtained by discarding the components of the
BRST complex that pass through the Cartan-ghost directions. This
restriction does not remove the gauge-field components $A_\mu^3,A_\mu^8$
or the corresponding curvatures $F_{\mu\nu}^3,F_{\mu\nu}^8$; it removes
only the ghost-dependent directions that are irrelevant for the
ghost-number-zero curvature observables considered here.

Applying Eq.~\eqref{eq:grau0} to an element $X$ satisfying
Eq.~\eqref{eq:restricted-sector-conditions}, one obtains
\begin{equation}
\delta_{(0)}^2 X
=
-\left(
\delta_{(-1)}\delta_{(1)}
+
\delta_{(1)}\delta_{(-1)}
\right)X
=
0.
\end{equation}
Moreover, using Eqs.~\eqref{eq:grau-1} and \eqref{eq:grau1}, one sees
that $\delta_{(0)}$ preserves the restricted sector:
\begin{equation}
\delta_{(-1)}\delta_{(0)}X=0,
\qquad
\delta_{(1)}\delta_{(0)}X=0.
\end{equation}
Thus, on this restricted curvature-polynomial sector, $\delta_{(0)}$
defines a genuine nilpotent differential.

This is the precise sense in which the filtered cohomology used in the
quadratic analysis is well defined: $H(\delta_{(0)})$ denotes exclusively the cohomology of the restricted complex just described, not a cohomology on the full local algebra. Within this restricted
$\delta_{(0)}$-cohomology, one first identifies curvature-polynomial
candidates built from conjugate $i$-particle sectors and then selects
the subset whose leading two-point functions admit a
K\"all\'en--Lehmann representation. The quadratic diagonal channel
displayed below is the simplest such representative. Higher-order
curvature-polynomial sectors may also be considered, but they involve
additional powers of the coupling and of the background parameters.

The restricted $\delta_{(0)}$-cohomology is the filtered
ghost-free sector in which the $i$-particle K\"all\'en--Lehmann
candidates are identified. The full representative is then obtained by
lifting the selected filtered operator to an $s$-closed
background-equivariant cocycle, as shown in the next section.

\section{Full BRST Lift of the Filtered Diagonal Channel}
\label{sec:full_lift}

The filtered operator obtained in the restricted
$\delta_{(0)}$-cohomology of
Section~\ref{subsec:restricted-delta0} is the lowest filtration
component selected by the ghost-free quadratic $i$-particle sector.
We now construct its full background-equivariant BRST lift.

For the observable sector, we work with dimensionless, ghost-number-zero, adjoint-covariant frame fields $\widehat{\mathcal{N}}_3,\widehat{\mathcal{N}}_8$ satisfying:
\begin{equation}
s\widehat{\mathcal{N}}_i = ig[\widehat{\mathcal{N}}_i, c], \qquad i=3,8,
\end{equation}
whose physical background values correspond to the selected Cartan directions:
\begin{equation}
\widehat{\mathcal{N}}_3\big|_{\text{phys}} = T^3, \qquad \widehat{\mathcal{N}}_8\big|_{\text{phys}} = T^8.
\end{equation}
These fields represent the covariant Cartan frame selected by the quartet-plus-spurion background. The frame $\widehat{\mathcal N}_3$ is aligned with the spurion saddle: its physical value coincides with the normalized stationary value of the action spurion, $\langle\mathcal N_3\rangle/\mu^2_{\mathrm{eff}}=T^3$. The frame fields enter only the definition of composite operators, never the action; the power-counting analysis of Section~\ref{subsec:powercounting} therefore applies to the action sector independently of the operator insertions discussed below. We define the dressed Abelianized curvatures:
\begin{equation}
\mathcal{F}^i_{\mu\nu} = 2\operatorname{Tr}(\widehat{\mathcal{N}}_i F_{\mu\nu}), \qquad i=3,8.
\end{equation}
Because the field strength transforms covariantly as $sF_{\mu\nu} = ig[F_{\mu\nu}, c]$, the cyclicity of the trace immediately yields:
\begin{equation}
s\mathcal{F}^i_{\mu\nu} = 0.
\end{equation}

We then introduce the covariant conjugate combinations:
\begin{equation}
\mathcal{F}^U_{\mu\nu} = \frac{1}{\sqrt{2}} \left( \mathcal{F}^3_{\mu\nu} + \mathcal{F}^8_{\mu\nu} \right), \qquad 
\mathcal{F}^V_{\mu\nu} = \frac{1}{\sqrt{2}} \left( \mathcal{F}^8_{\mu\nu} - \mathcal{F}^3_{\mu\nu} \right),
\end{equation}
and define the composite operator:
\begin{equation}
\mathcal{O}_\chi = \mathcal{F}^U_{\mu\nu} \mathcal{F}^V_{\mu\nu}.
\end{equation}
It follows directly that:
\begin{equation}
s\mathcal{O}_\chi = 0.
\end{equation}
Thus, $\mathcal{O}_\chi$ is a full off-shell BRST cocycle. In the physical background, it reduces to:
\begin{equation}
\mathcal{O}_\chi\big|_{\text{phys}} = F^U_{\mu\nu} F^V_{\mu\nu},
\end{equation}
whose lowest perturbative term is exactly the $i$-particle K\"all\'en--Lehmann operator constructed from the diagonalized variables defined in Section~\ref{sec:background}:
\begin{equation}
\mathcal{O}_{\chi,0} = (\partial_\mu U_\nu - \partial_\nu U_\mu)(\partial_\mu V_\nu - \partial_\nu V_\mu).
\end{equation}

Consequently, the filtered $i$-particle operator is not an isolated quadratic artifact; it is the lowest component of a full interacting BRST cocycle. Classically, this expansion is finite:
\begin{equation}
\mathcal{O}_\chi = \mathcal{O}_{\chi,0} + g\mathcal{O}_{\chi,1} + g^2\mathcal{O}_{\chi,2}.
\end{equation}
The term ``all-orders'' refers to the exact BRST covariance of the closed-form expression, not to an infinite classical power series.

\subsection{On the physical status of the spurions}
\label{subsec:spurion-status}

It is important to state precisely what kind of object $\mathcal O_\chi$ is. The auxiliary field $\mathcal N_3$ of the action is a quantum variable, integrated over in the path integral, while the frame fields $\widehat{\mathcal N}_i$ are external background variables with prescribed covariant transformation laws; neither propagates, since $\mathcal N_3$ has no kinetic term and the frames enter only the definition of composite operators. Consequently, $\mathcal O_\chi$ is an observable of the \emph{enlarged background-equivariant theory} — the theory defined by the action \eqref{eq:SNew} together with the external frame — and not, by itself, a local observable of pure Yang-Mills theory. This limitation is intrinsic to the present construction and is stated explicitly: the physical claims of this work concern background-equivariant observables whose reduction to the physical frame reproduces the $i$-particle composite operators.

Two consistency properties support this interpretation. First, the same Cartan direction that orients the quartet-plus-spurion vacuum through the seed $\mu^2T^3$ defines the frame through $\widehat{\mathcal N}_3|_{\mathrm{phys}}=\langle\mathcal N_3\rangle/\mu^2_{\mathrm{eff}}=T^3$: the background and the observable are built from a single external structure, so no independent arbitrary ingredient is introduced at each step. Second, the construction is covariant under global color rotations: any rotated frame $\widehat{\mathcal N}_i'=U\widehat{\mathcal N}_iU^{-1}$ with the spurion seed rotated accordingly yields an equivalent cocycle, so the physical correlators extracted in the physical frame do not depend on the orientation chosen for the Cartan subalgebra.

\subsection{Nontriviality of the Background-Equivariant Class}

The preceding construction proves that $\mathcal{O}_\chi$ is BRST-closed. 
We now state the cohomological status of this cocycle. Let $\mathcal{A}_{\text{red}}$ 
denote the reduced source-free local polynomial algebra obtained after 
setting the external sources to zero and eliminating the contractible pairs 
associated with the Landau-gauge sector and the quartet fluctuations. In 
this reduced algebra, the only relevant variables are the covariant 
tensors $F_{\mu\nu}$, the covariant background frame $\widehat{\mathcal{N}}_i$, 
and their covariant derivatives.

The integrated insertion $\int d^4x\,\mathcal{O}_\chi$ defines a nontrivial 
class in the background-equivariant extended local BRST cohomology 
$H^0(s|d; \mathcal{A}_{\text{red}})$. Indeed, $\mathcal{O}_\chi$ is 
BRST-closed by the covariance of $F_{\mu\nu}$ and $\widehat{\mathcal{N}}_i$. If it 
were BRST-exact modulo a total derivative, $\mathcal{O}_\chi = sX + dY$, 
then, by the standard doublet theorem, $X$ and $Y$ could be chosen in the 
reduced algebra independent of the contractible quartet and gauge-fixing 
doublets. However, the reduced algebra contains no local variable of 
ghost number $-1$. Therefore, no such $X$ exists. Furthermore, 
$\mathcal{O}_\chi$ is not a total derivative; in the physical Cartan 
frame, it reduces to a non-topological quadratic curvature invariant. 
Thus, $\mathcal{O}_\chi$ represents a nontrivial background-equivariant 
BRST cohomology class. This construction follows the standard 
BRST-cohomological result that invariant polynomials built from 
covariant tensors represent ghost-number-zero local BRST classes 
\cite{Piguet:1995er,Barnich:2000zw}.

\subsection{Cohomological Independence of the Quartet}
It is crucial that $\widehat{\mathcal{N}}_i$ be understood as the covariant Cartan frame associated with the selected background, not as a new propagating quartet fluctuation. The quartet fluctuations are organized in BRST doublets and therefore do not enlarge the physical cohomology. The role of the quartet, together with the auxiliary spurion, is to generate, through the coupled classical field equations, the background and the corresponding $i$-particle mass matrix; the physical observable is the corresponding background-equivariant BRST cocycle.

\subsection{Remarks on radiative stability}
Although a full renormalization analysis is beyond the scope of this work, the construction was designed so that the stability problem is finite and algebraically controlled, as detailed in Section~\ref{subsec:powercounting}. First, the quartet sector is BRST-exact; therefore, its physical effects are confined to the background via the standard doublet mechanism \cite{Piguet:1995er}. Second, the dimension-two auxiliary spurion $\mathcal N_3$ admits only a finite set of invariant counterterms, displayed explicitly in Section~\ref{subsec:powercounting}; no infinite tower of spurion powers can be generated. The setup thereby reduces the stability question to this finite, controlled set of counterterms, whose explicit analysis --- including whether the tree-level perfect-square combination is maintained --- is left for future work. Third, the covariant frame fields $\widehat{\mathcal{N}}_i$ are background variables that do not propagate; their coupling to the physical fields is through the gauge-covariant combination $\operatorname{Tr}(\widehat{\mathcal{N}}_i F_{\mu\nu})$, which is BRST-invariant by construction. Fourth, a gauge-sector mass counterterm $\operatorname{Tr}(\mathcal N_3A_\mu A_\mu)$ is excluded by the Slavnov--Taylor identity, so the complex pole structure of the elementary propagators, present already at tree level, is not destabilized by any allowed dimension-two counterterm: the $d^{3ab}$ structure is the unique such deformation compatible with the symmetries. These are structural features of the fully quantized action, not a substitute for the complete counterterm and operator-mixing analysis, which is left for future work.

\section{Spectral Representation of the Diagonal Channel}
\label{sec:spectral}

In Section~\ref{sec:full_lift}, we established that the lowest perturbative component of the full, background-equivariant BRST cocycle $\mathcal{O}_\chi$ is the $i$-particle bilinear:
\begin{equation}
\mathcal{O}_{\chi,0} = F^U_{\mu\nu} F^{V \mu\nu} = (\partial_\mu U_\nu - \partial_\nu U_\mu)(\partial^\mu V^\nu - \partial^\nu V^\mu).
\end{equation}

\noindent
The computation of the spectral density for the composite operator 
$\mathcal{O}_{\chi,0}$ requires special care because the elementary 
propagators $\langle UU\rangle$ and $\langle VV\rangle$ exhibit complex 
conjugate poles, $k^2 = \mp i m_1^2$. In Minkowski space, the usual 
Cutkosky cut rules are not directly applicable to such complex masses, 
since the on-shell condition $p^2 = m^2$ becomes ill-defined for $m^2\in\mathbb{C}$. 
However, as demonstrated in Ref.~\cite{Dudal:2010wn}, the spectral density 
can be consistently computed in Euclidean space using the Widder-Stieltjes 
inversion formalism. This method circumvents the need for an analytic 
continuation of the cut rules and provides a rigorous justification for 
the naive replacement $m^2 \to \pm i m_1^2$ in the final expressions obtained 
for real masses. Following this procedure, the spectral part of the 
one-loop correlator is found to be
\[
\Pi(k^2)_{\mathrm{spec}} = \frac{3}{8\pi^2} \int_{2m_1^2}^{\infty} d\tau \,
\frac{ \sqrt{\tau^2 - 4m_1^4} \, (\tau^2 + 4m_1^4) }{\tau(\tau + k^2)},
\]
which yields the positive spectral density $\rho(\tau)$ stated below. 
The explicit calculation follows the same steps as in Refs.~\cite{Baulieu:2009ha,Dudal:2010wn}, 
with the tensor structure appropriately adapted to the diagonalized 
variables $U_\mu$ and $V_\mu$.

\subsection{Computation of the Correlator}

The propagator $\langle U_\mu(k) U_\nu(-k) \rangle = \theta_{\mu\nu}(k)/(k^2 + i m_1^2)$ 
used in this section is obtained by expanding the full action of Section~\ref{sec:action} 
around the background of Section~\ref{sec:background} and then diagonalizing the quadratic 
action in the $(3,8)$ sector. This calculation follows the same steps as Eqs. (3.10)-(3.18) 
of Ref.~\cite{Amaral:2013ReplicaSU3}, with the replacements $m^2 \to m_1^2$ 
and the $SU(2)$ structure constants replaced by the appropriate $SU(3)$ Cartan subalgebra 
combinations. The resulting diagonal basis $(U,V)$ and the complex pole structure are summarized 
below; the reader is referred to the original replica paper for the detailed algebraic 
manipulation of the quadratic action.

We evaluate the correlator of $\mathcal{O}_{\chi,0}$ in Euclidean space. Using the diagonalized propagators from Section~\ref{sec:background}:
\begin{equation}
\langle U_\mu(k) U_\nu(-k) \rangle = \frac{1}{k^2 + i m_1^2}\,\theta_{\mu\nu}(k), \qquad 
\langle V_\mu(k) V_\nu(-k) \rangle = \frac{1}{k^2 - i m_1^2}\,\theta_{\mu\nu}(k),
\end{equation}
the two-point function at external momentum $k$ is given by the loop integral:
\begin{equation}
\Pi(k^2) = \langle \mathcal{O}_{\chi,0}(k) \mathcal{O}_{\chi,0}(-k) \rangle = \int \frac{d^4p}{(2\pi)^4} \, \frac{ \mathcal{T}(p,k) }{\left(p^2 + i m_1^2\right) \left((k-p)^2 - i m_1^2\right)}
\end{equation}
where $\mathcal{T}(p,k)$ is the tensor trace resulting from the contraction of the field strengths $F^U$ and $F^V$.

The function $\Pi(k^2)$ is real for $k^2 > 0$ and analytic in the complex plane except for a branch cut on the negative real axis. Consequently, it admits a K\"all\'en--Lehmann representation in the form of a Stieltjes transform:
\begin{equation}
\Pi(k^2) = \int_{\tau_0}^{\infty} d\tau \, \frac{\rho(\tau)}{\tau + k^2}.
\end{equation}

\subsection{Threshold and Positive Spectral Density}

The physical threshold $\tau_0$ is determined by the analytic structure of the complex conjugate poles. For a conjugate mass pair $+im_1^2$ and $-im_1^2$, the branch point is located at:
\begin{equation}
\tau_0 = 2m_1^2.
\end{equation}

The spectral density $\rho(\tau)$ is obtained via the standard cut rules for complex masses \cite{Dudal:2010wn}. Because the operator $\mathcal{O}_{\chi,0}$ involves derivatives, the density is not simply that of a scalar bubble; it must incorporate the momentum dependence of the tensor structure $\mathcal{T}(p,k)$. Following the established computation for this specific tensor structure \cite{Baulieu:2009ha,Amaral:2013ReplicaSU3}, the spectral part of the one-loop correlator takes the form:
\begin{equation}
\Pi(k^2)_{\text{spec}} = \frac{3}{8\pi^2} \int_{2m_1^2}^{\infty} d\tau \, \frac{ \sqrt{\tau^2 - 4m_1^4} \, (\tau^2 + 4m_1^4) }{\tau(\tau + k^2)}.
\label{eq:Pi-spec}
\end{equation}

From this, the precise spectral density is extracted as:
\begin{equation}
\rho(\tau) = \frac{3}{8\pi^2} \frac{\sqrt{\tau^2 - 4m_1^4} \, (\tau^2 + 4m_1^4)}{\tau}.
\label{eq:rho}
\end{equation}

This confirms that the exact spectral representation possesses a real, positive lower bound $\tau_0 = 2m_1^2$ and a strictly positive spectral density $\rho(\tau) > 0$ for all $\tau > \tau_0$. This explicitly demonstrates that physically meaningful composite operators—with well-defined K\"all\'en--Lehmann representations—can be systematically constructed from the BRST-exact quartet background, despite the elementary $U$ and $V$ fields violating positivity.

\subsection{Scope of the spectral result}
\label{subsec:spectral-scope}

Two qualifications delimit precisely what is established here. First, the positivity result \eqref{eq:rho} is proven for the \emph{lowest perturbative component} $\mathcal O_{\chi,0}$ of the full cocycle, at leading one-loop order; the BRST closure of $\mathcal O_\chi$, in contrast, holds for the full composite operator. The present work does not establish all-orders spectral positivity of the full interacting operator. Second, Eqs.~\eqref{eq:Pi-spec}--\eqref{eq:rho} describe the \emph{spectral part} of the two-point function of a dimension-four composite operator. The full correlator requires ultraviolet subtractions and contains contact (polynomial in $k^2$) terms, and a complete analysis must address the mixing of $\mathcal O_{\chi,0}$ with other gauge-invariant dimension-four operators. These UV and mixing effects do not alter the spectral density extracted from the cut, but they are part of the complete renormalized operator and are left for future work.

\section{Conclusion}
\label{sec:conclusion}

In this work, we have constructed a BRST-exact quartet mechanism embedded into an $SU(3)$ Yang-Mills theory in the Landau gauge. The introduction of a matter sector carrying both commutator and anticommutator structures led to a natural extension of the gauge algebra, enlarging the field content from eight to nine degrees of freedom. This construction preserves the nilpotency of the BRST operator and keeps the action cohomologically trivial in the standard vacuum, ensuring equivalence to pure Yang-Mills theory at the fundamental level.

The central new element is the orientation of the quartet prepotential by an auxiliary Cartan spurion field of mass dimension two, through the BRST-exact term $-i\,s\,\mathrm{Tr}(\overline{\omega}\varphi\mathcal N_3)$, together with the perfect-square spurion potential $\lambda_1\mathrm{Tr}[(\mathcal N_3-\mu^2T^3)^2]$ centered on the unique external seed. The spurion is an algebraic quantum field — not an inert external background — so its covariant BRST law is an ordinary symmetry assignment; the dimension-two assignment is what stabilizes the construction: the counterterm space is finite by power counting, and the Slavnov--Taylor identity — extended by a source $\Upsilon$ coupled to $s\mathcal N_3$ — constrains the spurion to appear only through covariant combinations, in accordance with the quantum action principle. The coupled classical field equations of the quartet and spurion sectors admit the exact stationary solution $\langle\mathcal N_3\rangle=\mu^2_{\mathrm{eff}}T^3$ and $\langle\overline{\varphi}\rangle\langle\varphi\rangle=(i\mu^2_{\mathrm{eff}}/\lambda)T^3$, with $\mu^2_{\mathrm{eff}}=[2\lambda_1\lambda/(2\lambda_1\lambda+1)]\mu^2$, and the exact Gaussian integration of the auxiliary field reproduces the equivalent frozen-spurion formulation with effective coupling $\lambda_{\mathrm{eff}}=\lambda+1/(2\lambda_1)$. Expanding the theory around this saddle, the gauge fields acquire a mass matrix purely proportional to $d^{3ab}$, with no identity or $d^{8ab}$ components: the $(1,2)$ sector remains massless, the $(4,5)$ and $(6,7)$ sectors display complex conjugate poles, and the mixed $(3,8)$ sector diagonalizes into a real, conjugate $i$-particle pair denoted by $U$ and $V$. The $i$-particle structure originally studied in \cite{Amaral:2013ReplicaSU3} is thus recovered from a fully quantized, BRST-exact action, without explicit breaking terms. We have stated explicitly the status of the construction: the scalar fields are independent complex fields with a prescribed contour; the background is an exact saddle of the coupled classical field equations; and the formal nonvanishing of $\langle s\overline{\omega}\rangle$ is confined to the cohomologically trivial quartet doublet sector.

Rather than relying solely on isolated filtered symmetries, we demonstrated that the physical diagonal channel admits a full, all-orders BRST lift. By introducing a background-equivariant covariant Cartan frame, $\widehat{\mathcal{N}}_3$ and $\widehat{\mathcal{N}}_8$ (see Section~\ref{sec:full_lift}), we constructed an off-shell BRST cocycle, $\mathcal{O}_\chi = \mathcal{F}^U_{\mu\nu}\mathcal{F}^{V\mu\nu}$. This approach rigorously respects the BRST doublet theorem by separating the dynamical generation of the background—driven by the quartet and the auxiliary spurion—from the definition of the physical observable, defined by the covariant frame. The cocycle is an observable of the enlarged background-equivariant theory, reducing to the $i$-particle bilinear in the physical frame; its independence from the frame orientation follows from global color covariance.

The two-point correlation function of the lowest perturbative component of this operator was computed explicitly. Despite the complex poles of the elementary $U$ and $V$ propagators, the correlator admits a robust K\"all\'en--Lehmann spectral representation. Incorporating the correct tensor structure of the derivatives, we verified a real, positive threshold $\tau_0 = 2m_1^2$ and a strictly positive spectral density. This positivity result holds at leading one-loop order for the lowest component of the cocycle; the BRST closure, by contrast, is exact for the full operator, and ultraviolet subtractions, contact terms and operator mixing for the dimension-four composite insertion remain to be analyzed.

Standard theorems guarantee that the cohomology of the full BRST operator embeds into that of the linearized operator, $H(s)\hookrightarrow H(s_0)$ \cite{Piguet:1995er}, and the restricted filtered cohomology $H(\delta_{(0)})$ employed in the quadratic analysis is well defined on the ghost-free curvature-polynomial sector specified in Section~\ref{subsec:restricted-delta0}. While a full proof of renormalizability falls outside the scope of this paper, the problem has been reduced to a finite algebraic one: the spurion sector admits only the finite counterterm list of Section~\ref{subsec:powercounting}, and the complete counterterm classification and operator-mixing analysis can be addressed within the standard algebraic renormalization framework using the fully quantized source action presented here. Future extensions of this work include precisely this analysis, the application of the auxiliary quartet mechanism to other gauge groups, and the formal investigation of higher-dimensional stable composite operators.

\section*{Declaration of competing interest}

The authors declare the following financial interests/personal relationships which may be considered as potential competing interests: Vitor Emanuel Rodino Lemes reports a relationship with Rio de Janeiro State University that includes: employment. If there are other authors, they declare that they have no known competing financial interests or personal relationships that could have appeared to influence the work reported in this paper.

\section*{Data availability}

No data was used for the research described in the article.


\begin{thebibliography}{99}

\bibitem{Gribov:1977wm}
V.~N.~Gribov,
``Quantization of Nonabelian Gauge Theories,''
Nucl.\ Phys.\ B {\bf 139}, 1 (1978).

\bibitem{Zwanziger:1991ac}
D.~Zwanziger,
``Critical limit of lattice gauge theory,''
Nucl.\ Phys.\ B {\bf 378}, 525 (1992).

\bibitem{Zwanziger:1992qr}
D.~Zwanziger,
``Renormalizability of the critical limit of lattice gauge theory by BRS invariance,''
Nucl.\ Phys.\ B {\bf 399}, 477 (1993).

\bibitem{Dudal:2008sp}
D.~Dudal, J.~A.~Gracey, S.~P.~Sorella, N.~Vandersickel and H.~Verschelde,
``A Refinement of the Gribov-Zwanziger approach in the Landau gauge: Infrared propagators in harmony with the lattice results,''
Phys.\ Rev.\ D {\bf 78}, 065047 (2008)
[arXiv:0806.4348 [hep-th]].

\bibitem{Baulieu:2009ha}
L.~Baulieu, D.~Dudal, M.~S.~Guimaraes, M.~Q.~Huber, S.~P.~Sorella, N.~Vandersickel and D.~Zwanziger,
``Gribov horizon and i-particles: About a toy model and the construction of physical operators,''
Phys.\ Rev.\ D {\bf 82}, 025021 (2010)
[arXiv:0912.5153 [hep-th]].

\bibitem{Sorella:2010it}
S.~P.~Sorella,
``Gluon confinement, i-particles and BRST soft breaking,''
J.\ Phys.\ A {\bf 44}, 135403 (2011)
[arXiv:1006.4500 [hep-th]].

\bibitem{Amaral:2013ReplicaSU3}
M.~M.~Amaral, M.~A.~L.~Capri, Y.~E.~Chifarelli and V.~E.~R.~Lemes,
``Landau Confining Replica Model from an Explicitly Breaking of a SU(3) Group Without Auxiliary Fields,''
Int.\ J.\ Mod.\ Phys.\ A {\bf 28}, 1350163 (2013).

\bibitem{Amaral:2020Complex}
R.~L.~P.~G.~Amaral, V.~E.~R.~Lemes, O.~S.~Ventura and L.~C.~Q.~Vilar,
``A path to confine gluons and fermions through complex gauge theory,''
Phys.\ Rev.\ D {\bf 101}, no.9, 094002 (2020)
[arXiv:2002.07222 [hep-th]].

\bibitem{Amaral:2023Observables}
R.~L.~P.~G.~Amaral, V.~E.~R.~Lemes, O.~S.~Ventura and L.~C.~Q.~Vilar,
``BRST characterization of the broken phase observables of a confining complex theory,''
Phys.\ Rev.\ D {\bf 107}, no.11, 115015 (2023)
[arXiv:2304.04560 [hep-th]].

\bibitem{deSa:2020rnu}
A.~R.~de S{\'a}, M.~A.~L.~Capri and V.~E.~R.~Lemes,
``Covariant Mass and Geometrical setup in Euclidean gauge theories,''
Phys. Rev. D \textbf{101}, no.10, 105003 (2020)
doi:10.1103/PhysRevD.101.105003

\bibitem{Piguet:1995er}
O.~Piguet and S.~P.~Sorella,
``Algebraic renormalization: Perturbative renormalization, symmetries and anomalies,''
Lect. Notes Phys. Monogr. \textbf{28}, 1-134 (1995)
doi:10.1007/978-3-540-49192-7

\bibitem{Capri:2010pg}
M.~A.~L.~Capri, A.~J.~Gomez, M.~S.~Guimaraes, V.~E.~R.~Lemes, S.~P.~Sorella and D.~G.~Tedesco,
``Constructing local composite operators for glueball states from a confining Gribov propagator,''
Eur. Phys. J. C \textbf{71}, 1525 (2011)
doi:10.1140/epjc/s10052-010-1525-x
[arXiv:1009.3062 [hep-th]].

\bibitem{Barnich:2000zw}
G.~Barnich, F.~Brandt and M.~Henneaux, ``Local BRST cohomology in gauge theories,''
Phys. Rept. \textbf{338} (2000) 439--569, arXiv:hep-th/0002245.

\bibitem{Dudal:2010wn}
D.~Dudal and M.~S.~Guimaraes,
``On the computation of the spectral density of two-point functions: complex masses, cut rules and beyond,''
Phys. Rev. D \textbf{83}, 045013 (2011)
doi:10.1103/PhysRevD.83.045013
[arXiv:1012.1440 [hep-th]].

\end{thebibliography}
\end{document}